\documentclass[12pt, draftclsnofoot, onecolumn]{IEEEtran}
\usepackage[caption=false,font=footnotesize]{subfig}
\usepackage{graphicx}
\usepackage{booktabs}
\usepackage{amsmath}
\usepackage{amssymb}
\usepackage{stfloats}
\usepackage{bm} 
\usepackage{color, cite}
\usepackage{algorithm}
\graphicspath{{figures/}}

\usepackage{enumerate,amsthm,}

\theoremstyle{plain}

\bstctlcite{BSTcontrol}

\begin{document}

\title{Interference Removal for Radar/Communication Co-existence: the Random Scattering Case}
\author{
Yinchuan Li, Le Zheng, \emph{Member}, \emph{IEEE}, Marco Lops, \emph{Fellow}, \emph{IEEE} and Xiaodong Wang, \emph{Fellow}, \emph{IEEE}
\thanks{

Y.~Li is with the School of Information and Electronics, Beijing Institute of Technology, Beijing 100081, China, the Beijing Key Laboratory of Embedded Real-time Information Processing Technology, Beijing 100081, China and Electrical Engineering Department, Columbia University, New York, USA, 10027 (e-mail: yinchuan.li.cn@gmail.com).

L.~Zheng and X.~Wang are with Electrical Engineering Department, Columbia University, New York, USA, 10027 (e-mail: le.zheng.cn@gmail.com; wangx@ee.columbia.edu).
		
M.~Lops is with the Department of Electrical Engineering and Information Technologies, Universit\`a di Napoli "Federico II", Via Claudio, 21 - I-80125 Naples (Italy) (e-mail: lops@unina.it).}
}
\maketitle

\begin{abstract}

In this paper we consider an un-cooperative spectrum sharing scenario, wherein a radar system is to be overlaid to a pre-existing wireless communication system. Given the order of magnitude of the transmitted powers in play, we focus on the issue of interference mitigation at the communication receiver. We explicitly account for the reverberation produced by the (typically high-power) radar transmitter whose signal hits scattering centers (whether targets or clutter) producing interference onto the communication receiver, which is assumed to operate in an un-synchronized and un-coordinated scenario. We first show that receiver design amounts to solving a non-convex problem of joint interference removal and data demodulation: next, we introduce two algorithms, both  exploiting sparsity of a proper representation of the interference and of the vector containing the errors of the data block. The first algorithm is basically a relaxed constrained  Atomic Norm minimization, while the latter relies on a two-stage processing structure and is based on alternating minimization. The merits of these algorithms are demonstrated through extensive simulations: interestingly, the two-stage alternating minimization algorithm turns out to achieve satisfactory performance with moderate computational complexity.

\end{abstract}
\begin{IEEEkeywords}
	Radar/communication co-existence, multi-path, atomic norm, compressed sensing, non-convex, blind deconvolution, off-grid, sparsity.
\end{IEEEkeywords}

\section{Introduction}

The ever increasing demand for high data rates in wireless communications has forced co-existence of communication and radar systems in the same frequency bands~\cite{griffiths2015radar}: this can be achieved by either allowing only one system to be equipped by an active properly designed transmitter - see, e.g. the information embedding strategies \cite{hassanien2016dual} to transmit information through a radar waveform, and the approach in \cite{Heath,802.11ad}, which can somehow be classified as a passive radar~\cite{griffiths2015radar}, to accomplish sensing functions through communication signals - or considering architectures with multiple transmitters operating in spectral overlap ~\cite{chiriyath2016inner,hessar2016spectrum,ding2016modified}. 

The latter scenario, which is the one considered in this paper, requires proper transceiver design: the strategies proposed so far range from a geometrical approach, aimed at mitigating the interference produced by one system on the other through suitable projection operations \cite{babaei2013practical,sodagari2012projection}, to a cognition-based radar waveform design \cite{deng2013interference,aubry2014radar, huang2015radar}. A more comprehensive approach is co-design \cite{lioptimum,zheng2018joint,turlapaty2014joint}, wherein the radar waveform(s) and the communication code-book are jointly designed by minimizing a measure of the mutual interference under certain constraints. A common point of these strategies is some form of coordination between the two active systems, and a remarkable degree of prior cognition, to be possibly acquired or updated through the periodic transmission of pilot signals to handle dynamic scenarios. 


In some situations, however, such a cooperation is either un-feasible - due, e.g., to security reasons - or too costly, whereby the radar and the communication systems should operate with little or no coordination. Such scenarios  have been considered, e.g., in \cite{khawar2014spectrum}, wherein a blind null space estimation method is proposed as an extension of the results of\cite{manolakos2012blind}. A different approach to handle un-coordinated co-existence is the one proposed in \cite{zheng2018adaptive}, considering full bandwidth overlap between a pre-existing communication system and multiple overlaid radars: assuming that the interfering radar waveforms live in the subspace of a known dictionary, the communication performance is guaranteed by joint interference removal/data demodulation iterative procedures, leveraging ideas from compressed sensing and atomic norm (AN) minimization techniques. A major limitation of \cite{zheng2018adaptive} is that the clutter induced by random scatterers disseminated in the controlled scene and reflecting the radar signal towards the communication receiver is not accounted for: this is a signal-dependent interference which, if not properly handled, typically produces dramatic effects on the radar performance and could totally prevent reliable communication. Additionally, synchronism between the radar and the communication system is assumed, as well as prior knowledge of the afore-mentioned dictionary. 

The present contribution is aimed at extending the results of \cite{zheng2018adaptive} by explicitly accounting for the reverberation produced by a single radar transmitter  onto the communication receiver. In particular, we consider an Orthogonal Frequency Division Multiplexing (OFDM) communication system co-existing with a short-range radar using a sophisticated waveform: the Pulse Repetition Interval (PRI) of the radar coincides with the duration of the communication data symbol block, and a totally un-synchronizated and un-coordinated scenario is considered, nor any assumption is made on the radar code structure. It is noteworthy that the PRI of the radar coincides with the duration of the communication data symbol block is possible in practice, as detailed in the next section. Similar to \cite{zheng2018adaptive}, we focus on the communication receiver performance, which is justified in the light of several considerations: first, the order of magnitudes of the powers transmitted by the communication and the radar transmitters is typically very different; additionally, while the communication transmitter points at the communication receiver whose location is typically known, whereby its effect on the radar receiver can be mitigated through beam-forming techniques\cite{liu2014joint}, a search radar employs rather wide and rotating beams, which produce random and time-varying reverberation onto the communication receiver. 
To this end, we propose two different algorithms, both exploiting two types of sparsity: on one hand, indeed, as scatterers are sparsely distributed in space, the interfering signals hitting the communication RX are {\em sparse}; on the other, an iterative demodulation algorithm should require that the vector containing the demodulation errors of a data block be itself sparser and sparser as the iterations go. Since the delays with which the interferers arrive at the communication receiver are continuous parameters, mere application of compressed sensing theory ~\cite{candes2011compressed,zhang2018recovery} would produce unsatisfactory performance ~\cite{chi2011sensitivity} in a situation where 
these signals cannot be sparsely represented by a finite discrete dictionary~\cite{stankovic2013compressive,jokanovic2015reduced,studer2012recovery}. We consider instead the recently developed mathematical theory of continuous sparse recovery for super-resolution~\cite{candes2013super,candes2014towards,tang2013compressed}, and especially of the AN minimization techniques which are successfully used for continuous frequency recovery, line spectral estimation and direction-of-arrival estimation~\cite{tang2013compressed,bhaskar2013atomic,tan2014direction}. As an alternative, based on the fact that
 the radar code is unknown and the radar interferences impinge on the communication RX with unknown multiple delays and coupling coefficients, estimating the interfering code and the multiple delays is inherently linked to solving a blind deconvolution problem~\cite{ling2017blind,ahmed2014blind,chi2016guaranteed}, which is non-convex and ill-posed without further constraints: this motivated us to also explore the recently developed mathematical theory of blind deconvolution~\cite{zhang2017global, zhang2018structured, chi2016guaranteed} to improve the estimation accuracy of the interfering waveform.


The remainder of this paper is organized as follows. In Section II, we present the signal models of the co-existed radar and communication system and set up the problem. In Section III, we develop the proposed convex relaxation method using both the AN and the $\ell_1$-norm. In Section IV, the proposed two-stage alternating minimization algorithm is developed. Simulation results are presented in Section V. Finally, in Section VI, we draw conclusions from the results obtained in this paper.

\begin{figure}[t]
	\centering

	\subfloat{\includegraphics[width=3.2in]{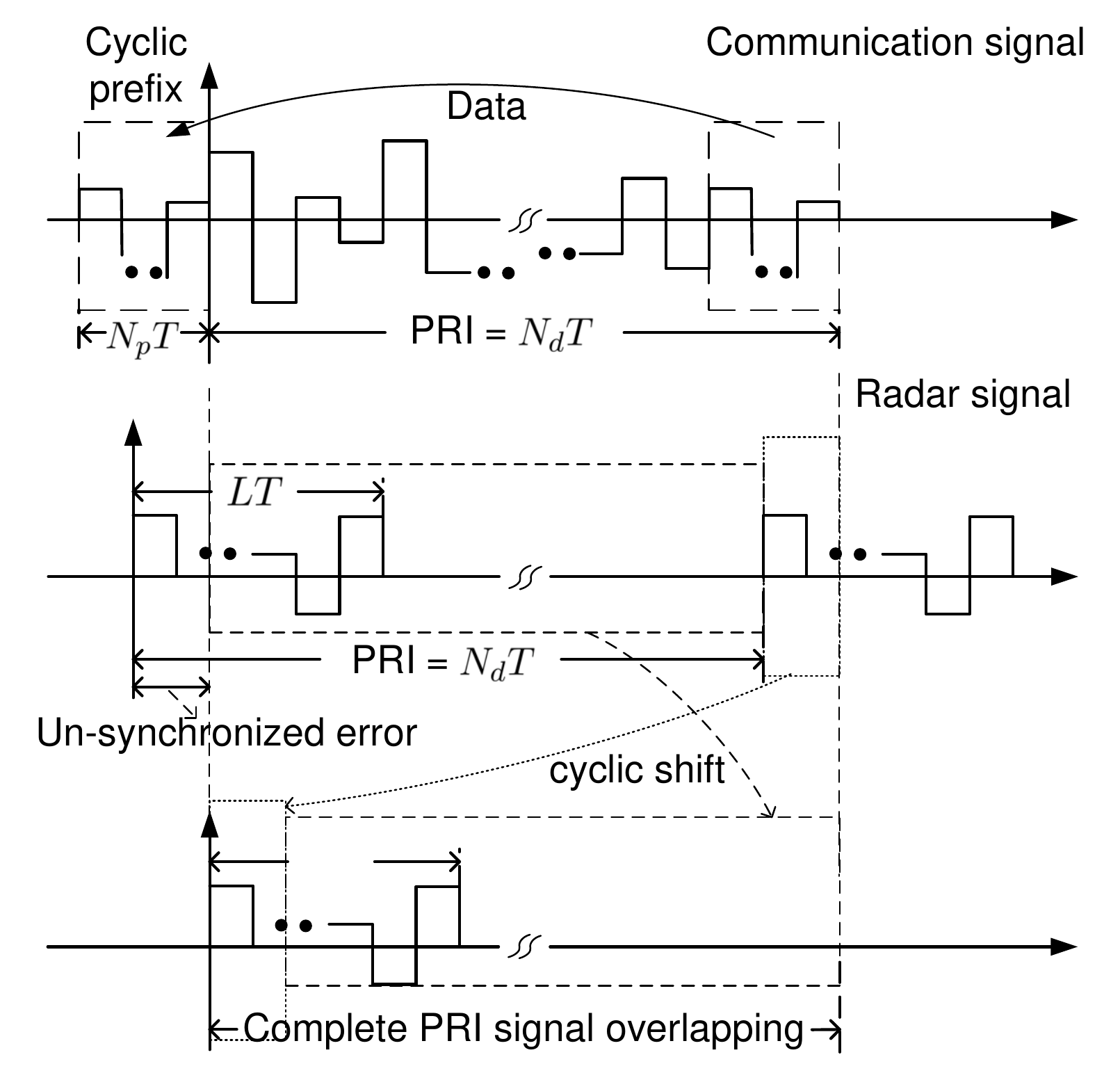}}

	\caption{Transmitted radar and communication signals}
	\label{figure:pulse}
\end{figure}

\section{System Descriptions \& Problem Formulation}

\subsection{Transmitted Signals}

We consider an OFDM communication system coexisting with a radar system. Assume that the OFDM system consists of $N = N_d + N_p$ sub-carriers, with $N_d$ data sub-carriers and $N_p$ cyclic prefix (CP) sub-carriers. The duration of an OFDM block is $NT$, $T$ being the ``sub-pulse duration.'' Denote the $n_c$-th normalized data symbol block as $b_{n_c}(k), k = 0,...,N_d-1$, such that $\mathbb{E}[b_{n_c}(k)b_{n_c}(k)^*]=1$ with $(\cdot)^*$ denoting the complex conjugate operator. Then, the transmitted baseband OFDM signal is given by
\begin{equation}
\label{eq:sc0}
s_c(t)= \sum_{n_c= -\infty}^{\infty} \sum_{k=0}^{N_d-1} b_{n_c}(k) e^{ i2\pi k \frac{t}{N_d T}} u_c(t - {n_c NT}),
\end{equation}
where 
\begin{equation}
\label{eq:xi0}
 u_c(t) = \left\{
\begin{aligned}
1, &~~~  t \in [-N_p T,N_d T], \\
0, &~~~  \text{otherwise}.
\end{aligned}
\right.
\end{equation}

As for the radar signal, we assume that the communication and the radar systems are in {\em full bandwidth overlap}, and the PRI equals the duration of data symbols $N_d T${\footnote{This assumption is possible in practice. For example, according to the December 2017 3GPP first release of the 5G New Radio standard, the data symbols of the 5G signal have a duration on the order of $10\mu s$, while short-range civilian radars (e.g., automotive radar) typically have a PRI in the order of $10\mu s$~\cite{engels2017advances}. In addition, some WLAN systems use OFDM waveform with a data symbol duration on the order of $1\mu s$~\cite{doufexi2002comparison}, while short-range impulse radars for high speed moving targets detection and through-the-wall radars may have a PRI in the order of $1\mu s$~\cite{zhiguo2006moving,wei2008detection}.}}. This assumption implies that, as shown in Fig.~\ref{figure:pulse}, the sub-pulse duration of the radar system and that of the communication system are the same; and at each PRI, a block of $N_d$ data symbols are transmitted. On the other hand, when the communication receiver processes one OFDM block of $N_d$ symbols, there is a complete PRI radar signal overlapping therewith after the cyclic shift, regardless of whether or not the two systems are synchronized.


We assume that the radar transmits a single {\em sophisticated} (i.e., with large duration-bandwidth product) pulse in any given PRI, which consists of $L$ amplitude-modulated sub-pulses. Denoting by $\bm g = [g(0),g(1),...,g(L-1)]^T  \in \mathbb{C}^{L \times 1}$ the waveform code and by $\xi(t)$ the basic sub-pulse waveform, the transmitted baseband radar signal is given by
\begin{eqnarray}
\label{eq:s_r}
s_r(t) = \sum_{n_r = -\infty}^{\infty}\sum_{\ell=0}^{L-1} g(\ell) \xi(t - \ell T - n_rN_d T),
\end{eqnarray}
where $T<<\text{PRI}$ is approximately the inverse of the bandwidth and is related to the radar range resolution. We remind here that the duty cycle $\delta=\frac{LT}{\text{PRI}}\ll1$ is typically low in order to guarantee a proper hearing period~\cite{wei2008detection}. 

\subsection{Received Signal}
We assume that the communication system operates on a block-fading channel  whose coherence time is much larger than the OFDM blocklength, whereby the useful component at the communication receiver is given by
\begin{align}
\label{eq:yc0}
y_c(t) = s_c(t) \ast h(t) = s_c(t) \ast \sum_{m = 1}^{M_c} \alpha_m \delta(t - \tau_m^c),
\end{align}
In the previous equation $\ast$ denotes the convolution operator, $h(t)$ is the channel impulse response, $M_c$ is the total number of propagation paths, $\alpha_m$ and $\tau_m^c$ are the $m$-th path's complex gain and delay, respectively.

The presence of a co-existing radar system produces additional interference on the communication receiver. In particular, if we assume that there are $M_r$ scatterers, whether from clutter or targets, located in as many different range cells, the signal scattered towards the communication receiver can be modeled
as
\begin{eqnarray}
\label{eq:yS0}
y_r(t) = s_r(t) \ast \sum_{m=1}^{M_r} c_m e^{j 2\pi f_m t} \delta(t - \tau_R - \tau_{m}^r),
\end{eqnarray}
where, since the radar and communication systems are un-synchronized, $0 \leq \tau_R \leq N_d T$ is the corresponding delay at a reference interval (i.e., for $n_c=n_r=0$), while $c_m$, $\tau_{m}^r$ and $f_m$ denote the scattering coefficient, the delay and the Doppler shift of the $m$-th reflector, respectively.

On the receiver side, we assume that the communication receiver processes one OFDM block of $N_d$ symbols at a time. Since the duration of an OFDM block is usually small, we have $f_mN_d T \ll 1$, then the phase rotation due to the Doppler shift over a block duration can be approximated as constant~\cite{berger2010signal}, and is thus not measurable and uninfluential, hence it is ignored from now on. The CP is removed assuming that its length is no less than the maximum communication multi-path delay. Let $n_c=0$ with no loss of generality and thus the subscript $n_c$ is also omitted. Focusing the attention on the interval $[0,N_d T]$, we thus have, for the received signal, the model:
\begin{align}
\label{eq:y0}
r(t) =& \sum_{n_r=-\infty}^{\infty}  \sum_{m=1}^{M_r} c_m \sum_{\ell=0}^{L-1} g(\ell)\xi(t- \ell T - n_r N_d T - \tau_R  - \tau_{m}^r)  \nonumber \\
&+  \sum_{m = 1}^{M_c} \alpha_m \sum_{k=0}^{N_d-1} b(k) e^{ i2\pi k \frac{t - \tau_m^c}{N_d T} } + \tilde w(t),~t\in[0,N_d T],
\end{align}
where $\tilde{w}(t)$ is a white, complex circularly symmetric Gaussian noise process.

\subsection{Problem Formulation}

The communication receiver is assumed to undertake the standard OFDM operations on each OFDM packet of duration $N_dT$. In particular, we focus on the first packet occupying the interval $[0,N_d T]$~\cite{molisch2012wireless}. Let
\begin{align}
\Delta_m \triangleq&~ \left\lfloor \frac{-\tau_R-\tau_m^r}{N_d T}\right\rfloor, \\
\tau_m \triangleq&~ \frac{-\tau_R-\tau_m^r}{N_d T} - \Delta_m  \in [0,1),
\end{align}
where $\lfloor \cdot \rfloor$ is the floor function. We have $\bar r(k)$ for $k = 0,...,N_d-1$, 
\begin{align}
\bar r(k) =&~ \frac{1}{N_dT} \int_{0}^{N_d T} {r(t) {e^{\frac{{ - i2\pi k t}}{N_dT}}}} dt \nonumber \\
=&~  \sum\limits_{m=1}^{M_r}  c_m \sum\limits_{\ell = 0}^{L-1}  g(\ell)   { \frac{1}{N_dT} \int_{0}^{N_d T} \sum_{n_r=-\infty}^{\infty} \xi(t- \ell T - n_r N_d T + \tau_{m} N_d T + \Delta_m  N_d T) {e^{\frac{{ - i2\pi k t}}{{N_d T}}}} dt } \nonumber \\
&~+ \sum_{m = 1}^{M_c} \alpha_m \sum_{n=0}^{N_d-1} b(n) \frac{1}{N_dT} \int_{0}^{N_d T}  e^{ i2\pi n \frac{t - \tau_m^c}{N_d T}} {e^{\frac{{ - i2\pi k t}}{{N_dT}}}} dt + w(k) \nonumber \\
\label{eq:barr0}
=&~  \sum\limits_{m=1}^{M_r}  c_m e^{ i {2\pi k} \tau_m} \sum\limits_{\ell = 0}^{L -1}
g(\ell) {e^{\frac{{ - i2\pi k \ell}}{{N_d}}}} {  \underbrace{ \frac{1}{N_dT} \int_{-\tau_m N_d T -\ell T }^{N_d T-\tau_m N_d T -\ell T} \sum_{n_r=-\infty}^{\infty} \xi( t - ( n_r - \Delta_m  )  N_d T ) {e^{\frac{{ - i2\pi k t}}{N_d T}}} dt }_{\bar\xi(\omega)| \omega = \frac{2\pi k}{N_dT} } } \nonumber \\
&~+ \sum_{m = 1}^{M_c} \alpha_m e^{ - i2\pi k \frac{\tau_m^c}{N_d T}} \sum\limits_{n = 0}^{N_d -1} {b(n)  \frac{1}{N_dT}  \underbrace{ \int_{0}^{N_d T}  e^{ i2\pi (n-k) \frac{t}{N_d T}}  dt }_{N_dT \cdot \delta(n-k) } } + w(k) \\
\label{eq:barr}
=&~ \sum\limits_{m=1}^{M_r} {c_m}{ {e^{ i2\pi k \tau_m}} {  \underbrace{\sum\limits_{\ell = 0}^{L -1}  g(\ell) {e^{\frac{{ - i2\pi k \ell}}{{N_d}}}} }_{\bar g(k)} { \bar\xi \left(\frac{2\pi k}{N_dT}\right) } } } + \underbrace{ \sum_{m = 1}^{M_c} \alpha_m e^{ - i2\pi k \frac{\tau_m^c}{N_d T}} }_{H(k)} b(k)   +  w(k),
\end{align}
where we note that in \eqref{eq:barr0} $\sum_{n_r=-\infty}^{\infty} \xi( t - ( n_r - \Delta_m  )  N_d T ) $ is a periodic signal with period $N_dT$, and each period is composed of $\xi(t), t\in[0,T]$; therefore the first integral is its Fourier transform, i.e., 
\begin{eqnarray}
\bar\xi(\omega) = \frac{1}{N_dT} \int_{0}^{N_d T}  \xi( t ) e^{-i \omega t} dt
\end{eqnarray}
evaluated at $\omega = \frac{2\pi k}{N_dT}$. In \eqref{eq:barr}, 
\begin{eqnarray}
w(k) \triangleq \frac{1}{N_dT} \int_{0}^{N_d T} {\tilde w(t) {e^{\frac{{ - i2\pi k t}}{N_dT}}}} dt \sim {\cal CN}(0,\sigma_w^2);
\end{eqnarray}
\begin{eqnarray}
\bm{\bar g} = [\bar g(0),\bar g(1),...,\bar g(N_d-1)]^T = \bm{F}_{L} \bm g \in \mathbb{C}^{N_d \times 1}
\end{eqnarray}
is the discrete Fourier transform (DFT) of $\bm g$, with $\bm{F}_{L}$ denoting the first $L$ columns of the $N_d$-points DFT matrix $\bm F$; and 
\begin{eqnarray}
H(k) = \int_{0}^{N_d T} {h(t) {e^{\frac{{ - i2\pi k t}}{N_dT}}}} dt = { \sum_{m = 1}^{M_c} \alpha_m e^{ - i2\pi k \frac{\tau_m^c}{N_d T}} }
\end{eqnarray}
 is the channel transfer function at frequency $\frac{k}{N_dT}$, which can be estimated using pilot signals~\cite{hu2011efficient}.

Let us now define 
\begin{eqnarray}
\bm{\bar\xi} = [\bar\xi(0),\bar\xi(\frac{2\pi}{N_dT}),...,\bar\xi(\frac{2\pi (N_d-1)}{N_dT})]^T\in \mathbb{C}^{N_d \times 1}
\end{eqnarray}
and
\begin{eqnarray}
\bm H = {\rm diag}( [H(0),H(1),...,H(N_d-1)]^T )\in \mathbb{C}^{N_d \times N_d},
\end{eqnarray}
i.e., an $N_d\times N_d$ diagonal matrix with elements of $[H(0),H(1),...,H(N_d-1)]^T$ on the diagonal. We also introduce the vectors 
\begin{align}
\bm{\bar r}=&~[\bar r(0),\bar r(1),...,\bar r(N_d - 1)]^T \in \mathbb{C}^{N_d \times 1}, \\
\bm b =&~[b(0),b(1),...,b(N_d-1)]^T \in \mathbb{C}^{N_d \times 1},\\
\bm {w} =&~ \left[w(0),w(1),...,w(N_d - 1) \right]^T \in \mathbb{C}^{N_d \times1}
\end{align}
and
\begin{eqnarray}
\label{eq:nu1}
\bm{\nu}_{\tau} = \sum_{m=1}^{M_r} c_m \bm a(\tau_m) \in \mathbb{C}^{N_d \times 1},
\end{eqnarray}
with 
\begin{eqnarray}
\bm a(\tau) = [1, e^{i2 \pi \tau},  ...,e^{i2\pi (N_d - 1) \tau}]^T.
\end{eqnarray}
Then, \eqref{eq:barr} can be given the following compact vector form
\begin{eqnarray}
\label{eq:vector-r}
\bm{\bar r} = \bm H \bm b + \bm{\bar\xi} \odot (\bm{F}_{L} \bm g) \odot \bm{\nu}_{\tau} + \bm w,
\end{eqnarray}
where $\odot$ denotes the pointwise product. Assume that an estimate of the data symbols, $\bm{\hat b}$, is available by directly performing demodulation using $\bm{\bar r}$. We subtract the
demodulated data from $\bm{\bar r}$, to obtain
\begin{eqnarray}
\label{eq:z}
\bm z = \bm{\bar r} - \bm H \bm{\hat b} = \bm H \bm v + \bm{\bar\xi} \odot (\bm{F}_{L} \bm g) \odot \bm{\nu}_{\tau} + \bm {w},
\end{eqnarray}
where 
\begin{eqnarray}
\bm v = \bm b - \bm{\hat b} \in \mathbb{C}^{N_d \times 1}.
\end{eqnarray}

Our main problem is to estimate $\bm g$, $\bm{\nu}_{\tau}$ and $\bm v$ from the noisy measurements $\bm z$. To this end, we first notice that, in a realistic scenario, the number of scatterers is much lower than the number of OFDM symbols in a packet, i.e. $M_r \ll N_d$ in \eqref{eq:nu1} ; secondly, we want the demodulation error rate to be low, i.e. we want to force the vector $\bm v$ to have a small number of non-zero entries: both are sparsity conditions that we can exploit. Notice however that the delays $\tau_m$ in \eqref{eq:nu1} take on continuous values, whereby using traditional compressed sensing techniques would entail heavy losses due to the off-grid problem: as a consequence, we resort here to Atomic Norm (AN) minimization instead~\cite{tang2013compressed,yang2016super}. Conversely, the second type of sparsity simply results in a suitable constraint in the optimization problem. To be more precise, define the set of atoms  ${\cal A} = \left\{ \bm a(\tau): \tau \in [0,1) \right\}$. Then the $\ell_0$-atomic norm~\cite{yang2016enhancing} associated to $\bm{\nu}_{\tau}$ is given by
\begin{eqnarray}
\label{eq:XA1}
\| \bm{\nu}_{\tau} \|_{{\cal A},0} = \inf_{c_m \in \mathbb{C}, \tau_m \in [0,1)} \left\{ M: \bm{\nu}_{\tau} = \sum_{m=1}^{M} c_m \bm a(\tau_m) \right\}.
\end{eqnarray}
Our problem can be formulated as
\begin{align}
\label{eq:atomicnorm1}
(\bm{\hat g}, \bm{\hat \nu}_{\tau}, \bm{\hat v}) =&~\arg \mathop {\min }\limits_{\substack{ \bm g \in \mathbb{C}^{L \times 1}, \bm{\nu}_{\tau}\in \mathbb{C}^{N_d \times 1} \\ \bm v \in \mathbb{C}^{N_d \times 1} } }    \| \bm{\nu}_{\tau} \|_{{\cal A},0} + \lambda \| \bm v \|_0, \\
&~\text{s.t.}~\left \| \bm z - \bm H \bm v - \bm{\bar\xi} \odot (\bm{F}_{L} \bm g) \odot \bm{\nu}_{\tau} \right \|_2^2  \leq \epsilon, ~ \| \bm g \|_2 = 1, \nonumber
\end{align}
where $\lambda>0$ is a weight factor, $\epsilon>0$ is the error tolerance and $\|\bm v\|_0 \ll N_d$ denotes the $\ell_0$-norm of $\bm v$. For the case that the radar signal is strong, we can perform iterative demodulation and radar interference estimation: in each iteration, after solving \eqref{eq:atomicnorm1}, we make use of $\bm{\hat v}$ and the current $\bm{\hat b}$ to obtain a refined demodulation
\begin{eqnarray}
\label{eq:estbCS}
\bm{\tilde b} = \arg\min_{\bm b \in {\cal B}^{N_d}} \| \bm b - \bm{\hat b} -  \bm{\hat v} \|_2,
\end{eqnarray}
where ${\cal B}$ is the modulation symbol constellation set. Then we update $\bm z$ in \eqref{eq:z} by setting $\bm{\hat b} \leftarrow \bm{\tilde b}$ and solve \eqref{eq:atomicnorm1} again.

Note that in \eqref{eq:atomicnorm1} the objective function is non-convex since it involves the $\ell_0$-atomic norm and the $\ell_0$-norm. The first constraint is also non-convex, because $(\bm{F}_{L} \bm g) \odot \bm{\nu}_{\tau}$ is the DFT of the convolution $\bm g \circledast ( \bm{F}^{-1}\bm{\nu}_{\tau})$ with $\circledast$ the circular convolution operator, and it is known that the blind deconvolution problem is non-convex~\cite{chi2016guaranteed, ling2017blind, ahmed2014blind}.

\section{The Convex Relaxation Method}

Define $\bm D = {\rm diag}(\bm{\bar\xi}) \bm{F}_{L} \in \mathbb{C}^{N_d \times L}$, and let $\bm d_k^H \in \mathbb{C}^{1 \times L}$ be the $k$-th row of $\bm D$. Then, \eqref{eq:z} can be rewritten as 
\begin{align}
\label{eq:z2}
z(k) =&~ \bm e_k^T (\bm H \bm v) + \bm{d}_k^H \bm g \bm e_k^T \bm{\nu}_{\tau} + w(k) = \bm e_k^H (\bm H \bm v) + \bm{d}_k^H (\bm g \bm{\nu}_{\tau}^T) \bm e_k + w(k) \nonumber \\ 
=&~ \left\langle \bm H  \bm v, \bm e_k \right\rangle + \left\langle \bm g \bm{\nu}_{\tau}^T , \bm{d}_k \bm e_k^H \right\rangle + w(k), ~k = 0,...,N_d-1,
\end{align}
where $z(k)$ denotes the $k$-th element of $\bm z$, $\bm e_k$ is the $k$-th column of the $N_d \times N_d$ identity matrix and $\left\langle \bm X, \bm Y \right\rangle = \text{Tr}(\bm Y^H \bm X)$. Notice that the original problem in \eqref{eq:atomicnorm1} entails estimating $\bm g$ and $\bm{\nu}_{\tau}$ separately. In the new formulation, we are interested in estimating $\bm g \bm{\nu}_{\tau}^T = \bm g \sum_{m=1}^{M_r} c_m \bm a(\tau_m)^T$
instead. In particular, we relax
\eqref{eq:z2} by introducing  
\begin{eqnarray}
\bm{X} = \sum_{m=1}^{M_r} c_m \bm g_m \bm a(\tau_m)^T \in \mathbb{C}^{L \times N_d},
\end{eqnarray}
i.e., a mixture of atoms from the atom set 
\begin{eqnarray}
{\cal {\tilde A}} = \left\{ \bm g \bm a(\tau)^T: \tau \in [0,1), \|\bm g\|_2 = 1, \bm g \in \mathbb{C}^{L \times 1} \right\}.
\end{eqnarray}
as the quantity of interest.
Further, we replace the $\ell_0$-atomic norm and the $\ell_0$-norm in the objective function of \eqref{eq:atomicnorm1} by the $\ell_1$-atomic norm~\cite{yang2016super} and the $\ell_1$-norm, respectively. The $\ell_1$-atomic norm seeks the tightest convex relaxation of enforcing sparsity in the atom set ${\cal {\tilde A}}$, and is defined as 
\begin{align}
\label{eq:XA2}
\| \bm X \|_{{\cal {\tilde A}},1} =&~\inf \left\{ \eta>0: \bm X \in \eta \text{conv}({\cal {\tilde A}}) \right\} = \inf_{\substack{  c_m \in \mathbb{C}, \tau_m \in [0,1) \\ \|\bm g_m\|_2 = 1 }} \left\{ \sum_m |c_m|: \bm X = \sum_m c_m \bm g_m \bm a(\tau_m)^T \right\},
\end{align}
where $\text{conv}(\cdot)$ denotes the convex hull of the input atom set. It is known that \eqref{eq:XA2} has the following equivalent form~\cite{yang2016super}:
\begin{align}
\label{eq:atomic}
\| \bm{X} \|_{{\cal {\tilde A}},1} = \mathop {\inf}\limits_{\bm u \in \mathbb{C}^{N_d \times 1}, \bm T \in \mathbb{C}^{L \times L}} \left\{ \begin{array}{l}
\frac{1}{2N_d}{\rm{Tr}}({\rm{Toep}}(\bm u)) + \frac{1}{2} {\rm Tr}(\bm T),\\
{\rm s.t.} \left[ {\begin{array}{*{20}{c}}
	{{\rm{Toep}}(\bm u)}& \bm{X}^H\\
	{{\bm{X}}}& {\bm T}
	\end{array}} \right] \succeq 0
\end{array} \right\},
\end{align}
where $\bm u \in \mathbb{C}^{N_d \times 1}$ is a complex vector whose first entry is real, ${\rm Toep}(\bm u)$ denotes the $N_d \times N_d$ Hermitian Toeplitz matrix whose first column is $\bm u$, and $\bm T$ is a Hermitian $L \times L$ matrix. Equations \eqref{eq:XA2} and \eqref{eq:atomic} are related through the relationship 
\begin{align}
{\rm Toep}(\bm u) =&~ \sum_m |c_m| \bm a(\tau_m) \bm a(\tau_m)^H, \\
\bm T =&~ \sum_m |c_m| \bm g_m \bm g_m^H.
\end{align}
Finally, we can relax the original problem in \eqref{eq:atomicnorm1} to the following semi-definite programming (SDP)~\cite{li2018atomic, chi2016guaranteed, zheng2018adaptive}
\begin{align}
\label{eq:SDP0}
(\bm{\hat X}, \bm{\hat v}) =&~ \arg \min_{ \substack{ \bm{X} \in \mathbb{C}^{L \times N_d}, \bm T \in \mathbb{C}^{L \times L} \\ \bm u \in \mathbb{C}^{N_d \times 1}, \bm v \in \mathbb{C}^{N_d \times 1} } }  \frac{{\text{Tr}}\left( {\rm Toep}(\bm u) \right)}{2N_d} + \frac{\text{Tr}(\bm T)}{2} +  {\bar\lambda} \|\bm v\|_1,  \\
 &~  \text{s.t.}~\sum_{k=0}^{N_d - 1} {\left| z(k) - \left\langle \bm H  \bm v , \bm e_k  \right\rangle - \left\langle \bm X , \bm{d}_k \bm e_k^H \right\rangle \right|^{2}} \leq \epsilon, \nonumber \\
 &~~~~~~ \left[ {\begin{array}{*{20}{c}}
	{{\rm Toep}(\bm u)}& \bm{X}^H \\
	{{\bm{X}}}& \bm T
	\end{array}} \right] \succeq 0, \nonumber
\end{align}
where $\bar\lambda>0$ is a weight factor. Since problem \eqref{eq:SDP0} is convex, it can be solved with standard convex solvers, e.g., CVX~\cite{boyd2004convex}.

Once an estimate $\bm{\hat X}$ of $\bm X$ is obtained, estimates of the delays $\{\tau_m\}$ and of the radar code $\bm g$ can be obtained by either solving the dual problem of \eqref{eq:SDP0} as in \cite{zheng2018adaptive,chi2016guaranteed}, or using the MUltiple SIgnal Classifier (MUSIC) method as in \cite{naha2015determining}. Note that by relaxing the original problem of estimating $\bm g \bm{\nu}_{\tau}^T = \bm g \sum_{m=1}^{M_r} c_m \bm a(\tau_m)^T$ to estimating $\bm{X} = \sum_{m=1}^{M_r} c_m \bm g_m \bm a(\tau_m)^T$, we may obtain spurious scatterers in solving the relaxed problem. As an example, suppose that the true code is $\bm g = [\frac{1}{\sqrt{2}},\frac{1}{\sqrt{2}},0,...,0]^T \in \mathbb{C}^{L\times 1}$, and there are two scatters with 
\begin{eqnarray}
{\begin{array}{*{20}{l}}
	\tau_1 = \tau \in [\frac{1}{N_d},1-\frac{1}{N_d}), & c_1 = \sqrt{2}, \\
	\tau_2 = \tau - \frac{1}{N_d}, & c_2=\sqrt{2}.
\end{array}}
\end{eqnarray}
Then the following is a spurious solution to the relaxed problem: 
\begin{align}
\bm g_1 =&~ [\frac{1}{\sqrt{3}},\frac{1}{\sqrt{3}},\frac{1}{\sqrt{3}},0,...,0]^T \in \mathbb{C}^{L\times 1},\nonumber \\
\bm g_2 =&~ [0,0,1,0,...,0]^T \in \mathbb{C}^{L\times 1}, \\
\tau_1' =&~ \tau,~ c_1'=\sqrt{3},~ \tau_2'=\tau + \frac{1}{N_d},~ c_2'=1, \nonumber
\end{align}
in the sense that we have
\begin{align}
\left\langle \bm g \bm{\nu}_{\tau}^T , \bm{d}_k \bm e_k^H \right\rangle &= \bar\xi(k) \left( e^{i2\pi k \tau} + 2e^{i2\pi k (\tau-\frac{1}{N_d})}  + e^{i2\pi k (\tau-\frac{2}{N_d})} \right)  \nonumber \\
&= \left\langle \bm X , \bm{d}_k \bm e_k^H \right\rangle, ~~ k = 0,...,N_d-1,
\end{align}
where 
\begin{eqnarray}
\bm{\nu}_{\tau} &=& c_1\bm a(\tau_1) +  c_2\bm a(\tau_2), \nonumber \\
\bm X &=& c_1'\bm g_1'\bm a(\tau_1')^T +  c_2'\bm g_2'\bm a(\tau_2')^T.
\end{eqnarray}

\section{A Two-Stage Alternating Minimization Algorithm}

Here we propose a new method to solve the non-convex problem in \eqref{eq:atomicnorm1}. The basic idea is to alternatively solve with respect to (w.r.t) $\bm g$ and $(\bm{\nu}_{\tau}, \bm v)$; and in solving w.r.t. $\bm g$, we use the conjugate gradient search on Riemannian manifold; in solving w.r.t. $(\bm{\nu}_{\tau}, \bm v)$, we take the matching-pursuit and greedy demixing approach. Moreover, we solve the problem in \eqref{eq:atomicnorm1} twice in a two-stage fashion: the first stage obtains a local optimum and the second stage makes use of the first-stage solution in forming the initial condition and solves a higher dimensional problem that leads to an approximate global optimum.

\subsection{Stage 1 - Obtaining Local Optimum} 

We first obtain a locally optimal solution to the non-convex problem \eqref{eq:atomicnorm1} by iteratively solving w.r.t. $\bm g$ and $(\bm{\nu}_{\tau}, \bm v)$ as follows:
\begin{description}
\item[S-1:] Let $\bm{\hat g}$ be the estimate - available from the previous iteration - of $\bm g$, and define 
\begin{eqnarray}
\bm \Phi \triangleq {\rm diag}(\bm{\bar\xi} \odot (\bm{F}_{L} \bm{\hat g})).
\end{eqnarray}
Then the new estimates $(\bm{\hat{\nu}}_{\tau}, \bm {\hat v})$ can be obtained by solving the problem:
\begin{align}
\label{eq:step-1}
(\bm{\hat{\nu}}_{\tau}, \bm {\hat v}) = \arg\min_{ \substack{\bm{\nu}_{\tau}\in \mathbb{C}^{N_d \times 1}\\ \bm v \in \mathbb{C}^{N_d \times 1} } } { \| \bm{\nu}_{\tau}  \|_{{\cal A},0} + \lambda \| \bm v \|_0}, ~\text{s.t.}~ {\left\| \bm z - \bm H \bm v - \bm \Phi \bm{\nu}_{\tau} \right\|_2^2 } \leq \epsilon.
\end{align}
\item[S-2:] With the estimates $\bm{\hat \nu}_{\tau}$ and $\bm{\hat v}$ of the previous step, defining
\begin{eqnarray}
\bm{\bar z} &\triangleq&  \bm z  - \bm H \bm{\hat v}, \\
\bm W &\triangleq& {\rm diag}(\bm{\bar\xi} \odot \bm{\hat \nu}_{\tau})\bm{F}_{L}.
\end{eqnarray}
$\bm g$ can be easily updated by solving:
\begin{align}
\label{eq:step-2}
\bm{\hat g} = \arg \mathop {\min }\limits_{\bm g \in \mathbb{C}^{L \times 1}} \left \| \bm{\bar z} - \bm W \bm g \right \|_2^2 , ~\text{s.t.}~\| \bm g \|_2 = 1.
\end{align}
\end{description}
The above alternating minimization procedure can be initialized by a random radar code $\bm{\hat g}$. We next present the details of the two steps.

\subsubsection{Greedy-demixing for solving S-1}

Since $\bm{\nu}_{\tau} = \sum_{m=1}^{M_r} c_m \bm a(\tau_m)$, estimating $\bm{\nu}_{\tau}$ in \eqref{eq:step-1} implies estimating $M_r$ as well as two vectors $\bm c = \left[{c_{1}},{c_{2}},...,{c_{M_r}}\right]^T \in \mathbb{C}^{M_r}$ and $\bm{\tau} = \left[{\tau_1},{\tau_2},...,{\tau_{M_r}}\right]^T \in [0,1)^{M_r}$ in $\bm{\nu}_{\tau} = \bm{\Theta}(\bm{\tau}) \bm {c}$, where
\begin{eqnarray}
\label{eq:Theta0}
	\bm{\Theta}(\bm{\tau}) = [\bm a(\tau_1), \bm a(\tau_2),..., \bm a(\tau_{M_r})] \in \mathbb{C}^{N_d \times M_r}.
\end{eqnarray}
If the delays are on-grid, it is easy to estimate $\bm v$ and $\bm{\tau}$ in \eqref{eq:step-1} using an orthogonal matching pursuit (OMP) algorithm~\cite{tropp2007signal}. However, since the delays here are off-grid, step S-1 involves not only demixing $\bm v$ and $\bm{\nu}_{\tau}$, but also super-resolving the delays in $\bm{\nu}_{\tau}$. In the spirit of the matching-pursuit method~\cite{lagarias1998convergence} and the greedy-demixing approach of~\cite{fernandez2017demixing}, we adopt the following procedure for solving \eqref{eq:step-1} in step S-1.
\begin{description}
\item[S-1(a)] \textbf{Initialization:} Let $\cal{S}$ be the set of support of $\bm v$, and $\cal{T}$ be the set of delays, and initialize them as  $\cal{S} \leftarrow \emptyset$, $\cal{T} \leftarrow \emptyset$. Define a residual $\bm r_{\text{res}} \in \mathbb{C}^{N_d\times 1}$ and initialize it as $\bm r_{\text{res}} \leftarrow \bm z$.

\item[S-1(b)] \textbf{Selection:} Find the highest correlation with the current residual $\bm r_{\text{res}}$ and update either $\cal{S}$ or $\cal{T}$. In particular, compute
\begin{align}
\label {eq:kmax}
k^{\diamond} =&~ \arg\max_{k\in \{0,1,...,N_d-1\} } | \langle \bm H(:,k), \bm r_{\text{res}} \rangle|, \\
\label {eq:taumax}
\tau^{\diamond} =&~ \arg\max_{\tau \in [0,1)}|\langle \bm \Phi \bm a({\tau}), \bm r_{\text{res}} \rangle|.
\end{align}
If $\tilde\lambda | \langle \bm H(:,k^{\diamond}), \bm r_{\text{res}} \rangle| > |\langle \bm \Phi \bm a(\tau^{\diamond}), \bm r_{\text{res}} \rangle| $, then ${\mathcal{S}} \leftarrow \mathcal{S} \cup \{  k^{\diamond}  \}$ otherwise ${\mathcal{T}} \leftarrow \mathcal{T} \cup \{ \tau^{\diamond} \}$, where $\tilde \lambda$ is a weight factor. To compute \eqref{eq:taumax}, we first search over a fine grid on $[0,1)$ with $N_f>N_d$ points. Then we perform a local search around the best grid point $\tau^{\diamond}_{\text{grid}}$. In particular, it is shown in Appendix A that \eqref{eq:taumax} has the following equivalent form
\begin{eqnarray}
\label{eq:cost-tau}
	\tau^{\diamond} =\arg\min_{\tau \in [0,1)} \text{Tr}\{\bm {A}^{\perp}(\tau) \bm{R}_{\text{res}} \},
\end{eqnarray}
where 
\begin{align}
\bm{R}_{\text{res}} =&~ (\bm \Phi^{-1} \bm{ \bm r_{\text{res}} }) (\bm \Phi^{-1}\bm{\bm  r_{\text{res}} })^{H} \in \mathbb{C}^{N_d \times N_d},\\
\bm {A}^{\perp}(\tau) =&~ \bm I_{N_d} - \frac{1}{N_d}\bm a({\tau}) \bm a({\tau})^H \in \mathbb{C}^{N_d \times N_d}.
\end{align}
Problem \eqref{eq:cost-tau} can be solved using Newton's method as 
\begin{eqnarray}
\label{eq:tau00}
	{\tau}^{i+1} = {\tau}^{i} - \mu_i K(\tau^{i})^{-1} {p(\tau^{i})},~~ i=0,1,...
\end{eqnarray}
where ${\tau}^{0} = \tau^{\diamond}_{\text{grid}}$; $\mu_i$ is the step size which is chosen according to the backtracking line search~\cite{bertsekas1999nonlinear}, given in Appendix B; ${p(\tau)}$ and $K(\tau)$ are the gradient and Hessian, given respectively by~\cite{viberg1991detection,viberg1991sensor}
\begin{align}
\label{eq:gradient0}
	{p(\tau)} =&~ \nabla_{\tau}\left[\text{Tr}\{\bm {A}^{\perp}({\tau}) \bm {R}_{\text{res}}\}\right] = -2{\rm Re}\left\{  \frac{1}{N_d} \bm{a}^H ({\tau})\bm {R}_{\text{res}} \bm {A}^{\perp}({\tau}) \frac{\partial \bm a(\tau)}{\partial \tau} \right\} \in \mathbb{R}, \\
\label{eq:Hessian0}
	K(\tau) =&~ \nabla_{\tau}^2 \left[\text{Tr}\{\bm {A}^{\perp}({\tau}) \bm {R}_{\text{res}}\}\right] \nonumber\\
	\approx&~ 2{\rm Re}\left\{ \left( (\frac{\partial \bm a(\tau)}{\partial \tau})^H \bm {A}^{\perp}({\tau}) \frac{\partial \bm a(\tau)}{\partial \tau} \right) { \frac{\bm{a}({\tau})^H \bm {R}_{\text{res}} \bm{a}({\tau})}{N_d^2}  } \right\} \in \mathbb{R} ,
\end{align}
where 
\begin{align}
\frac{\partial \bm a(\tau)}{\partial \tau} = \left[1,i2\pi e^{i2\pi \tau},...,i2\pi(N_d-1) e^{i2\pi(N_d-1)\tau}\right]^T \in \mathbb{C}^{N_d \times 1}.
\end{align}
The iteration in \eqref{eq:tau00} stops when $|K({\tau}^i)^{-1} p({\tau}^i) | < \delta$, where $\delta$ is the error tolerance, or the maximum iteration number $I$ is reached.

\item[S-1(c)] \textbf{Updating $\bm\tau$ using Newton's method:} If $\cal{T}$ is updated in step S-1(b), for the current $\bm{\hat v}$, using $\bm{\bar z} = \bm z  - \bm H \bm{\hat v}$, we refine the estimates of the delays in $\cal{T}$ by solving the following problem
\begin{eqnarray}
\label{eq:nls0}
\min_{\bm{\tau} \in [0,1)^{|\cal{T}|}, \bm {c}  \in \mathbb{C}^{|\cal{T}|}} {\left\| \bm{\bar z} - \bm \Phi \bm{\Theta}(\bm{\tau}) \bm {c}  \right\|_2^2}.
\end{eqnarray}
Substituting the solution $\bm{\hat c} = \bm{\Theta}(\bm{\tau})^{ \dagger}  \bm \Phi^{-1}  \bm{\bar z}$, where $(\cdot)^{\dagger}$ denotes the pseudo-inverse, i.e., $\bm{Y}^{ \dagger} = (\bm{Y}^H \bm{Y})^{-1}\bm{Y}^H$, back to \eqref{eq:nls0} results in the following optimization problem~\cite{viberg1991detection,viberg1991sensor}:
\begin{eqnarray}
\label{eq:cost-function}
	\bm{\hat{\tau}} = \arg\min_{ \bm{\tau} \in [0,1)^{|\cal{T}|} } \text{Tr}\{\bm {P}^{\perp}(\bm{\tau}) \bm {R}\},
\end{eqnarray}
where 
\begin{align}
\bm {R} =&~ (\bm \Phi^{-1} \bm{\bar z}) (\bm \Phi^{-1}\bm{\bar z})^{H} \in \mathbb{C}^{N_d \times N_d}, \\
\bm {P}^{\perp}(\bm{\tau}) =&~ \bm I_{N_d} - \bm{\Theta}(\bm{\tau}) \bm{\Theta}(\bm{\tau})^{ \dagger}\in \mathbb{C}^{N_d \times N_d}.
\end{align}
Problem \eqref{eq:cost-function} can be solved using Newton's method as 
\begin{eqnarray}
\label{eq:tau0}
	\bm{\tau}^{i+1} = \bm{\tau}^{i} - \bar\mu_i \bm K(\bm\tau^{i})^{-1} {\bm p(\bm\tau^{i})},~~ i=0,1,...
\end{eqnarray}
where $\bm{\tau}^{0}$ is taken as the current elements in $\cal{T}$; $\bar\mu_i$ is the step size which is chosen according to the backtracking line search~\cite{bertsekas1999nonlinear}, given in Appendix B; ${\bm p(\bm\tau)}$ and $\bm K(\bm\tau)$ are the gradient and Hessian matrix, given respectively by~\cite{viberg1991detection,viberg1991sensor}
\begin{align}
\label{eq:gradient0}
	{\bm p(\bm\tau)} =&~ \nabla_{\tau}\left[\text{Tr}\{\bm {P}^{\perp}(\bm{\tau}) \bm {R}\}\right] = -2{\rm Re}\left\{ {\text{vec-diag}} \left[\bm{\Theta}^{\dagger} (\bm{\tau})\bm {R} \bm {P}^{\perp}(\bm{\tau}) \bm {T}(\bm\tau) \right]\right\} \in \mathbb{R}^{{|\cal{T}|} \times 1}, \\
\label{eq:Hessian0}
	\bm K(\bm\tau) =&~ \nabla_{\tau}^2 \left[\text{Tr}\{\bm {P}^{\perp}(\bm{\tau}) \bm {R}\}\right] \in \mathbb{R}^{{|\cal{T}|} \times {|\cal{T}|}} \nonumber\\
	\approx&~ 2{\rm Re}\left\{ (\bm {T}(\bm{\tau})^H \bm {P}^{\perp}(\bm{\tau}) \bm {T}(\bm{\tau})) \odot ({ \bm{\Theta}(\bm{\tau})^{\dagger} \bm {R} \bm{\Theta}(\bm{\tau})^{\dagger H} })^T \right\},
\end{align}
where ${\text{vec-diag}}[\bm Y]$, with $\bm Y$ being a square matrix, denotes a column vector formed by the diagonal elements of $\bm Y$, and $\bm {T}(\bm{\tau})$ is given by
\begin{align}
\label{eq:bmT}
	&\bm {T}(\bm{\tau}) = \left[ \frac{\partial \bm a(\tau)}{\partial \tau} {\bigg |}_{\tau=\tau_1}, \frac{\partial \bm a(\tau)}{\partial \tau} {\bigg |}_{\tau=\tau_2},..., \frac{\partial \bm a(\tau)}{\partial \tau} {\bigg |}_{\tau=\tau_{|\cal{T}|}} \right] \in \mathbb{C}^{N_d \times {|\cal{T}|}} \nonumber\\
	 =& \left[ {\begin{array}{*{20}{c}}
	1 & 1 & \cdots & 1 \\
	i2\pi e^{i2\pi \tau_1}& i2\pi e^{i2\pi \tau_2} & \cdots & i2\pi e^{i2\pi \tau_{|\cal{T}|}}\\
	\vdots & \vdots & \ddots & \vdots \\
	i2\pi(N_d-1) e^{i2\pi(N_d-1)\tau_1} & i2\pi(N_d-1) e^{i2\pi(N_d-1)\tau_2} & \cdots & i2\pi(N_d-1) e^{i2\pi(N_d-1)\tau_{|\cal{T}|}}
	\end{array}} \right].
\end{align}
The iteration in \eqref{eq:tau0} stops when $\|\bm K(\bm{\tau}^i)^{-1} \bm p(\bm{\tau}^i) \|_2 < \delta$ or the maximum iteration number $I$ is reached.

\item[S-1(d)] \textbf{Updating $(\bm v,\bm c)$ using least-squares:} With the current $\mathcal{S}$ and $\mathcal{T}$, estimate $\bm v$ and $\bm c$ by solving the following least-squares problem:
\begin{align}
\label {eq:least-squares}
(\bm{\hat v}, \bm{\hat c}) = \arg\min_{ \substack{ \bm v({\cal{S}}) \in \mathbb{C}^{|{\cal{S}}|}\\  \bm {c}  \in \mathbb{C}^{{|\cal{T}|}} } } {\left\| \bm z - \bm H(:,{\cal{S}}) \bm v({\cal{S}}) - \bm \Phi \bm{\Theta}(\bm{\hat{\tau}}) \bm {c}  \right\|_2^2},
\end{align}
where $\bm H(:,{\cal{S}})$ and $\bm v({\cal{S}})$ denote the columns and elements of $\bm H$ and $\bm v$ respectively indexed by ${\cal{S}}$. Then, we remove any atoms in $\mathcal{T}$ whose corresponding coefficients are smaller than a small threshold $\tilde\delta$.

\item[S-1(e)] \textbf{Residual update:} 
\begin{eqnarray}
\bm r_{\text{res}} = \bm z - \bm H \bm{\hat v} - \bm \Phi \bm{\Theta}(\bm{\hat{\tau}})  \bm{\hat c}
\end{eqnarray}
and repeat steps S-1(b) to S-1(e) until $\| \bm r_{\text{res}} \|_2^2 \leq \epsilon$, or the maximum iteration number $I'$ is reached.
\end{description}

\subsubsection{Conjugate gradient descent for solving S-2}

The constraint $\| \bm g \|_2 = 1$ in \eqref{eq:step-2} can be regarded as forcing $\bm g$ on a unit sphere, which belongs to the Riemannian manifolds. We thus resort to the conjugate gradient method on Riemannian manifold~\cite{chen2017low} to update $\bm g$. The Euclidean gradient of the cost function in \eqref{eq:step-2} is 
\begin{eqnarray}
\bm q(\bm g) = \nabla_{\bm g}\left \| \bm{\bar z} - \bm W \bm g \right \|_2^2 = -2\bm W^H(\bm{\bar z} - \bm W \bm g).
\end{eqnarray}
Projecting the Euclidean gradient to the tangent space of Riemannian manifold yields the Riemannian gradient~\cite{chen2017low,absil2009optimization}
\begin{eqnarray}
\bm q_{\text{R}}(\bm g) = \bm q(\bm g) - {\rm Re}(\bm q(\bm g) \odot \bm g^*)\odot \bm g.
\end{eqnarray}
Then the search direction in the $i$-th iteration is given by~\cite{chen2017low}
\begin{align}
\label{eq:conjugate-gradient}
\bm q_{\text{C}}(\bm g^i) = \gamma_i [ \bm q_{\text{C}}(\bm g^{i-1}) - {\rm Re} (\bm q_{\text{C}}(\bm g^{i-1}) \odot (\bm g^i)^*)\odot \bm g^i ] - \bm q_{\text{R}}(\bm g^i), ~~ i=1,2..., 
\end{align}
with $\bm q_{\text{C}}(\bm g^0) = - \bm q_{\text{R}}(\bm g^0)$, where
\begin{eqnarray}
\gamma_i = \frac{\bm q_{\text{R}}(\bm g^i)^T(\bm q_{\text{R}}(\bm g^i) - \bm q_{\text{R}}(\bm g^{i-1}))}{\|\bm q_{\text{R}}(\bm g^{i-1})\|_2^2}.
\end{eqnarray}
Then, problem \eqref{eq:step-2} can be solved by the following iterations
\begin{eqnarray}
\label{eq:g1}
\bm g^{i+1} = \frac{\bm g^i + \tilde\mu_i \bm q_{\text{C}}(\bm g^i)}{\|\bm g^i + \tilde\mu_i \bm q_{\text{C}}(\bm g^i)\|_2^2}, ~~ i=0,1,...,
\end{eqnarray}
where $\tilde\mu_i$ is a step size, which is also chosen according to the backtracking line search~\cite{bertsekas1999nonlinear}, given in Appendix~B. The iteration in \eqref{eq:g1} stops when $\|\bm g^{i+1} -\bm g^{i} \|_2 < \bar\delta$, where $\bar\delta$ is the error tolerance, or the maximum iteration number $\bar I$ is reached.

\subsection{Stage 2 - Inferring the Global Optimum}

\begin{figure}[!t]
	\centering
		
	\subfloat{\includegraphics[width=2.5in]{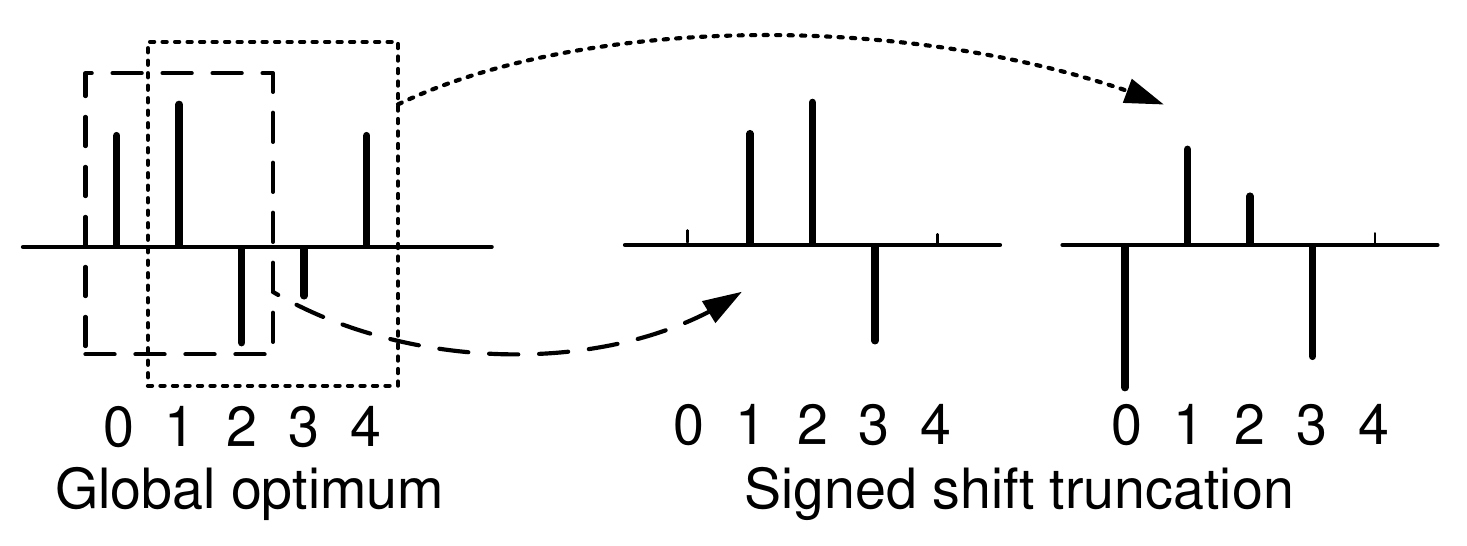}}
	
	\caption{Two examples of signed shift truncation.}
	\label{figure:signed}
\end{figure}

\begin{figure}[!t]
	\centering
		
	\subfloat{\includegraphics[width=2.8in]{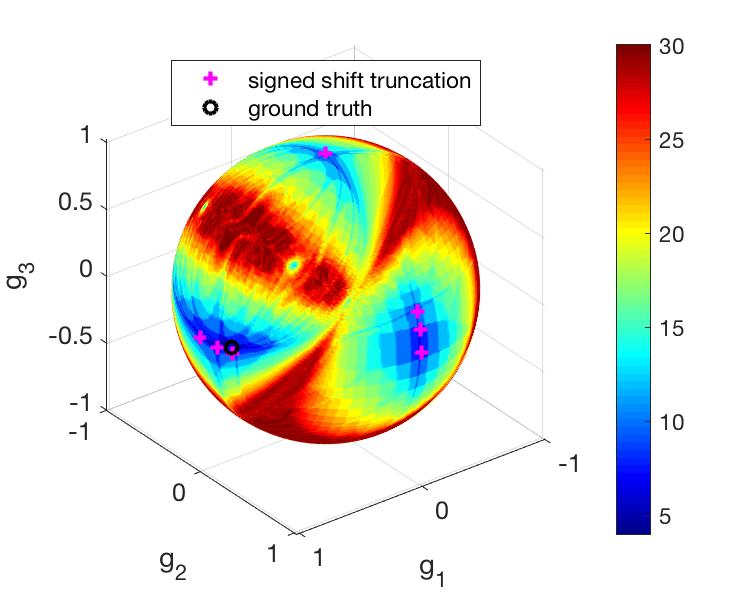}}
	
	\caption{Geometry of the objective function of \eqref{eq:atomicnorm1} on the $\ell^2$ ball when error $\bm v = 0$. Dark blue represents small values indicating a local optimum. All local optima are close to signed shift truncations of the ground truth $\bm g_{\star}$.}
	\label{fig:sphere}
\end{figure}

After Stage 1, we obtain a locally optimum solution $(\bm{\hat g},\bm{\hat\tau},\bm{\hat c})$ to problem \eqref{eq:atomicnorm1}. To further search for the global optimum, we make use of a theoretical result in~\cite{zhang2018structured}. Recall that $(\bm{F}_{L} \bm g) \odot \bm{\nu}_{\tau} = \bm F (\bm g \circledast (\bm F^{-1}\bm{\nu}_{\tau} ))$. When $\bm g \in \mathbb{C}^{L\times1}$ with $L\ll N_d$, and $\bm F^{-1}\bm{\nu}_{\tau} \in \mathbb{C}^{N_d}$ is a sparse vector, then $\bm g \circledast (\bm F^{-1}\bm{\nu}_{\tau} )$ is the so-called short-and-sparse (SaS) convolution. It is shown in~\cite{zhang2018structured} that, for the SaS blind deconvolution problem, if $\bm F^{-1}\bm{\nu}_{\tau}$ follows the Bernoulli-Gaussian (BG) model, then any local optimum $\bm{\hat g}$ is close to certain signed shift truncation of the global optimum $\bm g_{\star}$ with high probability. The signed shift truncation is the result of truncation, circular shift and sign changes on a sequence (see the two examples in Fig.~\ref{figure:signed}). Hence, we speculate that the estimated code $\bm{\hat g}$ obtained by Stage 1 is close to a signed shift truncation of the global optimum $\bm g_{\star}$. In fact, this conjecture is corroborated by extensive simulations. For example, the landscape of the objective function in \eqref{eq:atomicnorm1} when $\bm v = 0$ and $L = 3$ is shown in Fig.~\ref{fig:sphere}. In particular, for a given point on the sphere $\|\bm g\|_2=1$, we calculate the corresponding $\min \| \bm{\nu}_{\tau}  \|_{{\cal A},0}$ via steps S-1(a)-(e). Dark blue represents small values while dark red represents large values. The landscape clearly shows that \eqref{eq:atomicnorm1} is non-convex. Furthermore, we calculate all the signed shift truncations of the ground truth $\bm g_{\star}$ and mark them on the sphere. We can see that the local optima are very close to certain signed shift truncations of the ground truth.

\begin{figure}[!h]
	\centering
		
	\subfloat{\includegraphics[width=2.5in]{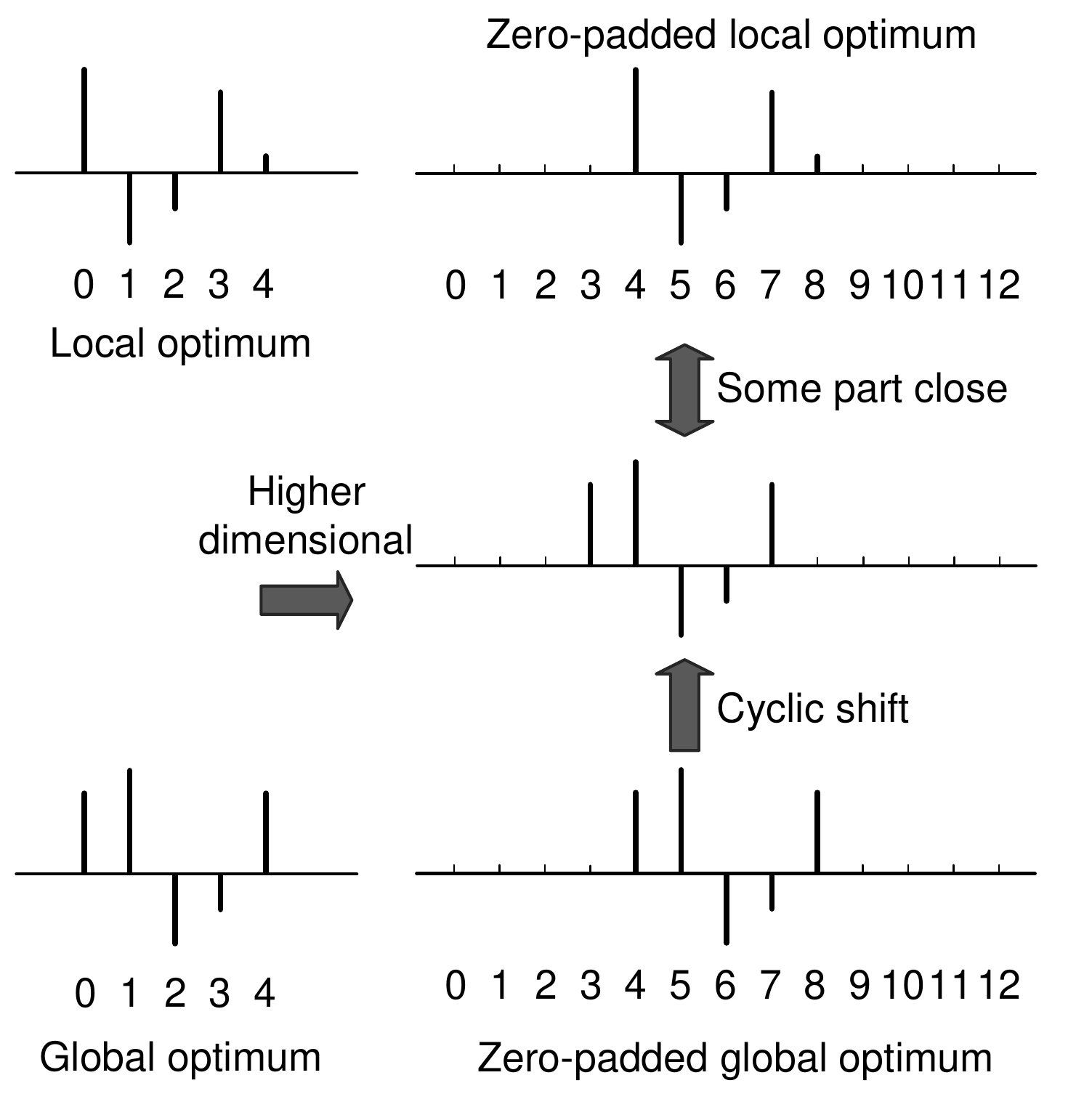}}
	
	\caption{An example of local optimum and global optimum, and their relationship in the higher dimensional space. Some part of the zero-padded local optimum and the cyclic shift of the zero-padded global optimum are close.}
	\label{figure:higher}
\end{figure}

As the local optimum still captures a considerable portion of the global optimum, then in a higher dimensional space, the zero-padded local optimum should be close to one cyclic shift of the zero-padded global optimum (see Fig.~\ref{figure:higher}). Hence, we first estimate a cyclic shifted zero-padded global optimum instead of estimating the global optimum directly, and the zero-padded local optimum serves as a significantly better initialization than a random initial value in a higher dimensional space~\cite{zhang2017global}.

\begin{figure}[!t]
	\centering
		
	\subfloat{\includegraphics[width=3.0in]{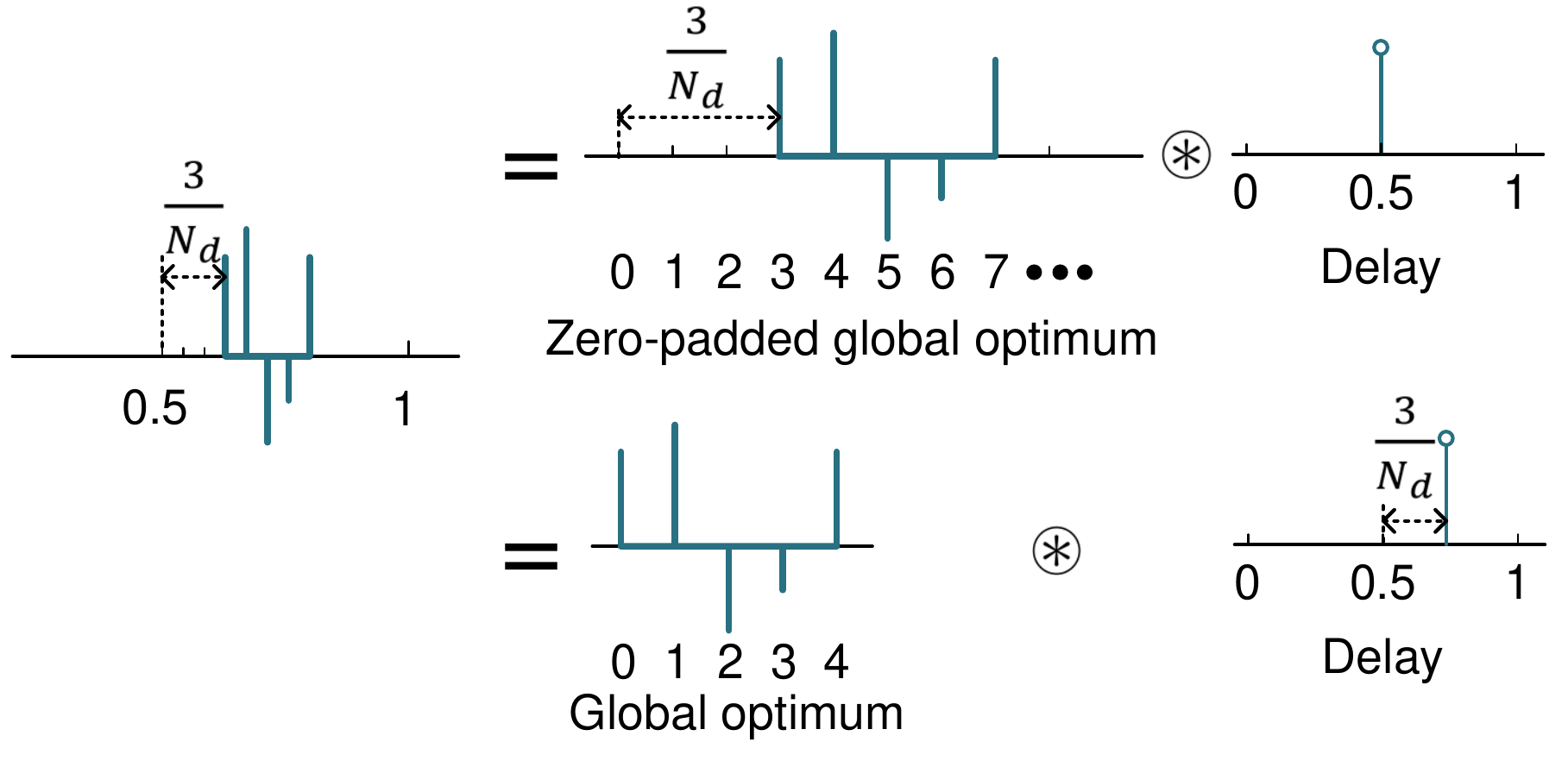}}
	
	\caption{Shift ambiguity in short-and-sparse deconvolution.}
	\label{figure:shift-delay}
\end{figure}

The estimated $\bm{\hat g}$ of Stage 1 after zero-padding is $\bm{\tilde g}_{0} = \left[ \bm {\bar{0}}^T, \bm{\hat g}^T, \bm { \bar{0}}^T \right]^T \in \mathbb{C}^{ (3L-2) \times 1}$, where $\bm {\bar{0}} \in \mathbb{C}^{ (L-1) \times 1}$ is the all-zero vector\footnote{Note that if $\bm{\hat g}$ is shifted by more than its own length $L$, there will be no truncation of global optimum $\bm g_{\star}$ retained. Therefore, the length of zero padding vector $\bm {\bar{0}}$ is set as $L-1$.}. Problem \eqref{eq:atomicnorm1} in a higher dimensional space is given by
\begin{align}
\label{eq:atomicnorm3}
(\bm{\tilde g}_{\text{h}}, \bm{\tilde \nu}, \bm{\tilde v}) =&~ \arg \mathop {\min }\limits_{ \substack{ \bm{\nu}\in \mathbb{C}^{N_d \times 1}, \bm v\in \mathbb{C}^{N_d \times 1} \\  \bm g_{\text{h}} \in \mathbb{C}^{(3L-2) \times 1} }  }    \| \bm{\nu} \|_{{\cal A},0} + \lambda \| \bm v \|_0, \\
&~\text{s.t.}~ \left \| \bm z - \bm H \bm v - \bm{\bar\xi} \odot (\bm{F}_{3L-2} \bm g_{\text{h}}) \odot \bm{\nu} \right \|_2^2  \leq \epsilon, ~\| \bm g_{\text{h}} \|_2 = 1, \nonumber
\end{align}
where $\bm{F}_{3L-2}$ denotes the first $3L-2$ columns of $\bm F$. Solving \eqref{eq:atomicnorm3} by using the same alternating minimization algorithm outlined in Section IV.A with initial value $\bm{\tilde g}_{0}$ yields the estimates $\bm {\tilde g}_{\text{h}}$, $\bm{\tilde c}$, $\bm{\tilde{\tau}}$ and $\bm{\tilde v}$. Since $\bm {\tilde g}_{\text{h}}$ is an estimate of the cyclic shifted zero-padded global optimum $\bm g_{\star}$, we need to extract the estimate of $\bm g_{\star}$ by detecting the first element that is larger than a small threshold $\delta_{\text{h}}$, i.e.,
\begin{align}
\label{eq:estb}
\tilde \ell =&~ \arg\min_{\ell \in {\{0,1,...,3L-2\}}} \ell , ~\text{s.t.} ~ \tilde g_{\text{h}}(\ell) > \delta_{\text{h}}, 
\end{align}
where $\tilde g_{\text{h}}(\ell)$ is the $\ell$-th element of $\bm{\tilde g}_{\text{h}}$. Then the estimated global optimum code is $\bm {\tilde g}_{\star} = [\tilde g_{\text{h}}(\tilde \ell), \tilde g_{\text{h}}(\tilde \ell+1),...,\tilde g_{\text{h}}(\tilde \ell+L-1)]^T \in \mathbb{C}^{L\times1}$. Note that convolution has the shift ambiguity property. Extracting $\bm g_{\star}$ from $\bm{\tilde{g}}_{\text{h}}$ is equivalent to a cyclic shift of length $-\frac{\tilde{\ell}}{N_d}$ on $\bm \tau$ in the convolution $\bm g \circledast (\bm F^{-1}\bm{\nu}_{\tau} )$ (see Fig.~\ref{figure:shift-delay}). Hence, the estimated global optimum delay is $\bm{\tilde\tau}_{\star} = \bm{\tilde{\tau}} + \frac{\tilde \ell}{N_d}$.

\begin{algorithm}[!t]\small
	\label{tab:A1}
	\caption{Two-stage alternating minimization procedure for solving \eqref{eq:atomicnorm1}.}
	\begin{tabular}{lcl}
		Input $\bm z$, $\bm H$, $\bm{\bar\xi}$, $\tilde\lambda$, $\epsilon$, $\delta$, $\bar\delta$, $\tilde\delta$, $I$, $I'$, $\bar I$ and $\delta_{\text{h}}$.\\
		1, Initialize $\bm{\hat g}$ as a random code.  \\
		{\sf{Repeat}} (Stage 1) \\
		\hspace{0.4cm} 2, Obtain $\bm{\hat v}$, $\bm{\hat\tau}$ and $\bm{\hat c}$ via S-1. \\
                 \hspace{0.4cm} 3, Obtain $\bm {\hat g}$ via S-2.\\
                 \hspace{0.4cm} 4, $\bm{\hat b} \leftarrow \bm{\tilde b} = \arg\min_{\bm b \in {\cal B}^{N_d}} \| \bm b - \bm{\hat b} -  \bm{\hat v} \|_2$.\\
                 \hspace{0.4cm} 5, $\bm z = \bm{\bar r} - \bm H \bm{\hat b}$.\\
                 \sf{Until} $\bm{\tilde b} = \bm{\hat b}$.\\
                 6, $\bm{\tilde g}_{0} = \left[ \bm {\bar{0}}^T, \bm{\hat g}^T, \bm { \bar{0}}^T \right]^T $.\\
                 {\sf{Repeat}} (Stage 2)\\
                 \hspace{0.4cm} 7, Obtain $\bm {\tilde g}_{\text{h}}$, $\bm{\tilde{\tau}}$, $\bm {\tilde c}$ and $\bm{\tilde v}$ by performing steps 2-5 with the \\
                 \hspace{0.75cm} higher dimensional $\bm{g}_{\text{h}}$ initialized as $\bm{\tilde g}_{0}$.\\
                 \sf{Until} $\bm{\tilde b} = \bm{\hat b}$.\\
                 8, Obtain $\tilde \ell$ by using \eqref{eq:estb}.\\
                 9, $\bm {\tilde g}_{\star} = [\tilde g_{\text{h}}(\tilde \ell), \tilde g_{\text{h}}(\tilde \ell+1),...,\tilde g_{\text{h}}(\tilde \ell+L-1)]^T$. \\
                 10, $\bm{\tilde\tau}_{\star} = \bm{\tilde{\tau}} + \frac{\tilde \ell}{N_d}$.\\
		\midrule
		Return $\bm {\tilde g}_{\star}$, $\bm{\tilde\tau}_{\star}$, $\bm{\tilde c}_{\star}$ and $\bm{\tilde v}_{\star}$.\\
	\end{tabular}
\end{algorithm}

Finally, we summarize the proposed two-stage alternating minimization (2-AltMin) method in Algorithm 1. The main computational load of Algorithm 1 is the calculation of gradient and Hessian in Newton's method, with complexities ${\cal O}(N_d^3)$ and ${\cal O}(M_rN_d^2)$ per iteration, respectively. Hence the computational complexity of Algorithm 1 is ${\cal O}(N_d^3)$ per iteration. On the other hand, the complexity of the convex relation (CR) method discussed in Section~III is ${\cal O}((N_d + L)^6)$ per iteration if the interior point method is used~\cite{zheng2018adaptive}. Hence the proposed 2-AltMin method is both computationally more efficient and more accurate as shown by simulation results in the next section.

\section{Simulation Results}

\subsection{Baseline for Comparison: On-grid Method}

As a baseline of comparison, we consider the on-grid method for estimating the continuous delays $\{\tau\}$, by using an overcomplete dictionary matrix
\begin{eqnarray}
\label{eq:grid-dictionary}
	\tilde{\bm A}= [\bm a_0, \bm a_1,..., \bm a_{\tilde M-1}] \in \mathbb{C}^{N_d \times \tilde M},
\end{eqnarray}
where $\tilde M \geq N_d$ and $\bm a_m = \bm a(\frac{m}{\tilde M}),~m=0,1,...,\tilde M-1$. For sufficiently large $\tilde M$, the delay is densely sampled. Following the convex relaxation used in Section III, define \begin{eqnarray}
\bm{\varsigma} = [c_{1} \bm{g}_1^T, c_{2} \bm{g}_2^T, ... , c_{\tilde M} \bm{g}_{\tilde M}^T ]^T \in \mathbb{C}^{\tilde ML \times 1}
\end{eqnarray}
as the sparse vector whose non-zero elements correspond to $c_m \bm g$ in \eqref{eq:z2}. The original problem \eqref{eq:atomicnorm1} can be relaxed to the following on-grid optimization problem
\begin{align}
\label{eq:cs}
(\bm{\hat{\varsigma}}, \bm{\hat v}) =&~ \arg \mathop {\min }\limits_{\substack{ \bm \varsigma \in \mathbb{C}^{\tilde ML \times 1}, \bm v \in \mathbb{C}^{N_d \times 1} } }    \| \bm{\varsigma} \|_1 + \bar\eta \| \bm v \|_1, \\
&~\text{s.t.}~ \left \| \bm z - \bm H \bm v - \bm \Upsilon \bm{\varsigma} \right \|_2^2  \leq \epsilon, \nonumber
\end{align}
where $\tilde \eta$ is a weight factor and $\bm \Upsilon$ is given by
\begin{align}
\label{eq:Upsilon}
\bm \Upsilon  = \left[ {\begin{array}{*{20}{c}}
	{\bm a_0^H {\bm e_{0}} \bm{d}_{0}^H}&{\bm a_1^H {\bm e_{0}} \bm{d}_{0}^H}& \cdots &{\bm a_{\tilde M-1}^H {\bm e_{0}} \bm{d}_{0}^H}\\
	{\bm a_0^H {\bm e_{1}} \bm{d}_{1}^H}&{\bm a_1^H {\bm e_{1}} \bm{d}_{1}^H}& \cdots &{\bm a_{\tilde M-1}^H {\bm e_{1}} \bm{d}_{1}^H}\\
	\vdots & \vdots & \ddots & \vdots \\
	{\bm a_0^H {\bm e_{N_d-1}} \bm{d}_{N_d-1}^H}&{\bm a_1^H {\bm e_{N_d-1}} \bm{d}_{N_d-1}^H}& \cdots &{\bm a_{\tilde M-1}^H {\bm e_{N_d-1}} \bm{d}_{N_d-1}^H}
	\end{array}} \right] \in \mathbb{C}^{N_d \times \tilde ML}.
\end{align}
Since problem \eqref{eq:cs} is convex, it can be solved with standard convex solvers, e.g., CVX~\cite{boyd2004convex}. And the complexity in each iteration is ${\cal O}((\tilde ML+N_d)^3)$ if the interior point method is used~\cite{zheng2018adaptive}. Then, the radar delays and code can be identified by locating the non-zero entries of $\bm{\hat{\varsigma}}$, i.e., if $[\hat\varsigma_{mL},\hat\varsigma_{mL+1},...,\hat\varsigma_{(m+1)L-1}]$ has elements larger than a pre-set small threshold, then a radar delay exists at $\frac{m}{\tilde M}N_dT$ and normalizing $[\hat\varsigma_{mL},\hat\varsigma_{mL+1},...,\hat\varsigma_{(m+1)L-1}]$ yields the corresponding estimated radar code.

Note that this on-grid method is also a relaxed method, and similar to the example given in Section III, it can be shown that some columns of $\bm \Upsilon$ can be identical. Hence, $\bm \Upsilon$ is {\em coherent}~\cite{candes2011compressed} and many delay false alarms could be generated in $\bm{\varsigma}$, which will be illustrated in the simulations.

\subsection{Simulation Setup}

In order to demonstrate the performance of the proposed algorithms, we consider a scenario where a radar transmitter produces multiple reflections towards a communication receiver. The communication system uses an OFDM signal with $N_d = 256$, $N_p = 64$ and a total bandwidth of $2.56$ MHz, i.e, the frequency spacing between adjacent subcarriers is $10$ kHz. Hence the duration of data symbols $N_dT = 100~\mu s$ and a quadrature phase-shift keying (QPSK) modulation is used. The transmitted OFDM signal is generated according to \eqref{eq:sc0} with normalized data symbols. Since the communication takes place over a multi-path Rayleigh-fading channel (see eq. \eqref{eq:yc0}), the path gains $\{\alpha_m\}$ are i.i.d. complex Gaussian distributed, $\alpha_m\sim {\cal CN}(0,\sigma_h^2)$. Based on \eqref{eq:barr}, we define the SNR at the communication RX as
\begin{align}
\text{SNR} = \frac{ \mathbb{E}\{| \sum_{m = 1}^{M_c} \alpha_m e^{ - i2\pi k \frac{\tau_m^c}{N_d T}} |^2 \} }{\sigma_w^2} = \frac{ \sum_{m = 1}^{M_c} \mathbb{E}\{|  \alpha_m |^2 \} }{\sigma_w^2} = \frac{ M_c \sigma_h^2 }{\sigma_w^2},
\end{align}
where $\sigma_w^2$ is the variance of the Gaussian noise sample $w(k)$ in \eqref{eq:barr}. In the following simulations, we set $M_c = 10$ and $\sigma_h^2=0.1$.

In \eqref{eq:s_r}, the radar uses a pulse coded waveform and pulse uses Gaussian random code with length $L$, and then we normalize the code to let $\|\bm g\|_2=1$ for the simplicity of evaluation. The sub-pulse of radar signal $\xi(t),t\in[0,T]$ is set as the normalized rectangular pulse of duration $T$. The reference delay $\tau_{R}$ and the delays $\tau_{m}^r$ of the scatters are randomly generated between $0$ and $100~\mu s$. The radar PRI is set as $N_d T = 100~\mu s$. The scatterers are modeled as point sources in our simulations, and the complex scattering coefficient $c_m$ of the $m$-th scatter is generated with fixed magnitude $c_0$ and random phase for convenience of evaluation. Specifically, based on \eqref{eq:barr}, we define the interference-to-signal ratio (ISR), which is the average power ratio of the radar interference and the communication signal, at the communication RX as
\begin{align}
\text{ISR}  = \frac{ \frac{1}{N_d} \sum\limits_{k=0}^{N_d-1} \mathbb{E} \left \{ \left | {\bar g(k)} { \bar\xi (\frac{2\pi k}{N_dT}) }\sum\limits_{m=1}^{M_r} {c_m}{ {e^{ i2\pi k \tau_m}} }  \right|^2 \right\} }{ M_c \sigma_h^2 } = \frac{ {{M_r} {|c_0|^2}} \sum\limits_{k=0}^{N_d-1} \left| {\bar g(k)} { \bar\xi (\frac{2\pi k}{N_dT}) }  \right|^2 }{ N_d M_c \sigma_h^2 }.
\end{align}

\begin{figure*}[!t]
	\centering
	
	\subfloat[][]{\includegraphics[width=3.2in]{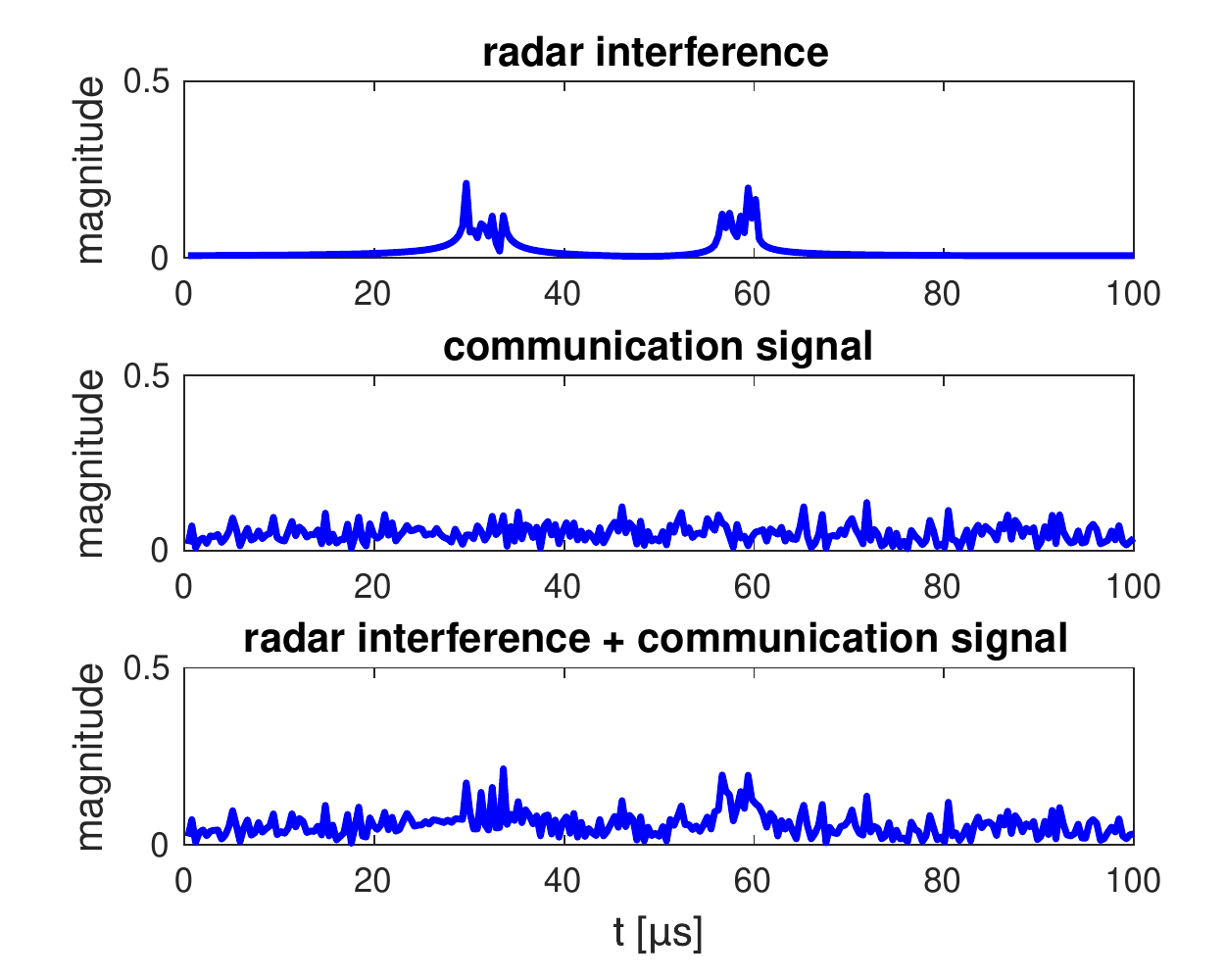}}
	\subfloat[][]{\includegraphics[width=3.2in]{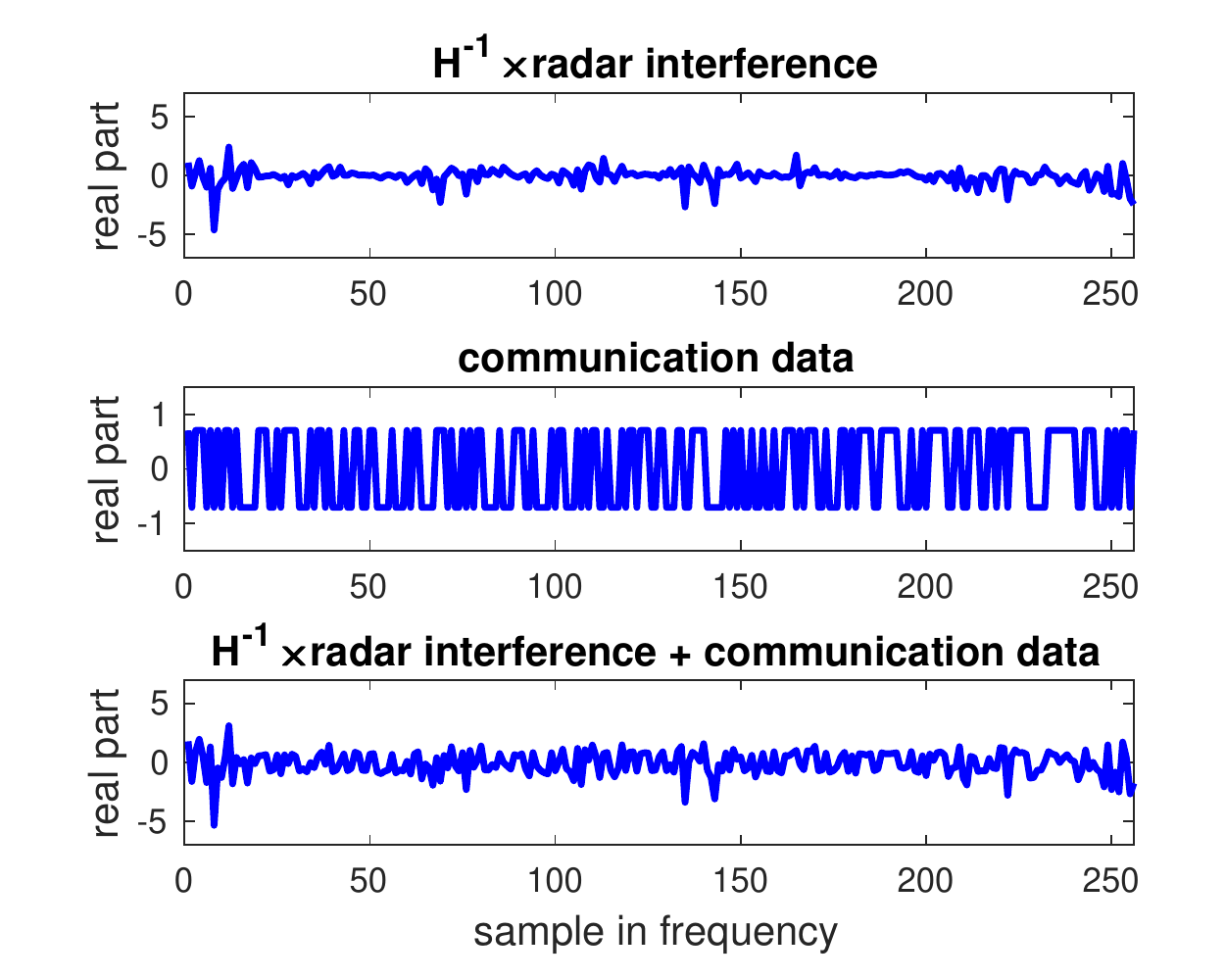}}
	
	\caption{Plots of signal waveforms in (a) time domain and (b) frequency domain. In (a), the magnitude of the radar interference, communication signal and the received signal of communication RX are plotted against time. In (b), the real part of the interference on communication data, communication data and their combination are plotted versus frequency sample.}
	\label{fig:example}
\end{figure*}

We evaluate the mean absolute error (MAE) of the radar delay estimate and the mean-squared-error (MSE)\footnote{We use the relative MSE rather than the RMSE to evaluate the accuracy because it reflects the loss in energy.} of the estimated radar code for the on-grid method, the CR method and the 2-AltMin algorithm. Note that in each case the algorithm returns a bunch of delays, which can be either true detections or false alarms. In the simulations, for each estimated delay $\hat\tau^r_{\ell}, {\ell} = 1,...,|{\cal{T}}|$, we calculate the  minimum absolute error $\text{AE}_{\ell}$ with the ground truth delays $\tau^r_{m},m=1,...,M_r$, i.e., $\text{AE}_{\ell} = \min(\{ \hat\tau^r_{\ell} - \tau^r_{m} \}_{m=1}^{M_r})$. Then, the delay MAE and the relative waveform MSE are respectively calculated as
\begin{align}
\label{eq:MAE}
\text{MAE}_{\tau} =&~ \frac{1}{\text{MC}} \sum_{n_{\text{MC}}=1}^{\text{MC}}  \frac{1}{|{\cal{T}}|}\sum_{\ell = 1}^{|{\cal{T}}|} \text{AE}_{\ell}^{(n_{\text{MC}})}, \\
\text{MSE}_{g} =&~ \frac{1}{\text{MC}} \sum_{n_{\text{MC}}=1}^{\text{MC}}  \frac{ \left\|  |{\bm g}^{(n_{\text{MC}})}| - |\bm {\tilde g}_{\star}^{(n_{\text{MC}})}| \right\|_2^2 } {\left\| |{\bm g}^{(n_{\text{MC}})}| \right\|_2^2} ,
\end{align}
where $\text{MC}$ is the number of Monte Carlo runs;  $\text{AE}_{\ell}^{(n_{\text{MC}})}$ is the minimum absolute error of the $\ell$-th estimate in the $n_{\text{MC}}$-th run; $\bm g^{(n_{\text{MC}})}$ and $\bm {\tilde g}_{\star}^{(n_{\text{MC}})}$ are the radar code and the estimated radar code at the $n_{\text{MC}}$-th run, respectively.

The error tolerance is usually set smaller than $\epsilon \backsimeq 0.05{\left\| \bm z \right\|_2^2}$, which implies that the iteration stops when the relative error is smaller than 5\%~\cite{li2012isar}. For the proposed algorithms, we set the error tolerance in \eqref{eq:atomicnorm1} as $\epsilon \backsimeq 0.01{\left\| \bm z \right\|_2^2}$ for better performance. The weight factors for the on-grid method in \eqref{eq:cs} and the CR method in \eqref{eq:SDP0} are respectively set as $\bar\eta = 1$ and $\bar\lambda = \frac{1}{\sqrt{N_d}}$~\cite{zheng2018adaptive}. And the weight factor for the 2-AltMin algorithm in step S-1(b) is set as $\tilde\lambda = {\frac{6}{\sqrt{N_d M_c \sigma_h^2}}}$. The grid number $\tilde M$ in \eqref{eq:grid-dictionary} is set as $\tilde M = 512$. The error tolerances for Newton's method and conjugate gradient method are both set as $\delta = \bar \delta = 10^{-6}$, and the threshold in step S-1(d) and \eqref{eq:estb} are respectively set as $\tilde\delta = 0.05$ and $\delta_{\text{h}} = 0.05$. The maximum iteration numbers for Newton's method, step S-1(e) and the conjugate gradient method are respectively set as $I = 10$, $I' = 50$ and $\bar I = 10$. In addition, ${\rho}$ and $\bar{\rho}$ for the backtracking line search in Algorithm 2 are respectively set as $0.5$ and $0.01$.

In order to show the performance of the proposed methods, we compare the symbol error rate (SER) of the proposed methods with the SER of directly performing demodulation using $\bm{\bar r}$, which is named ``Iteration 0" because its result is the initial point of the iterative algorithms.
In addition, we compare the performance of the Stage 1 of the 2-AltMin method, which is named ``Stage 1''.

\subsection{Performance}

\begin{figure*}[!htp]
	\centering
	
	\subfloat[][]{\includegraphics[width=3.2in]{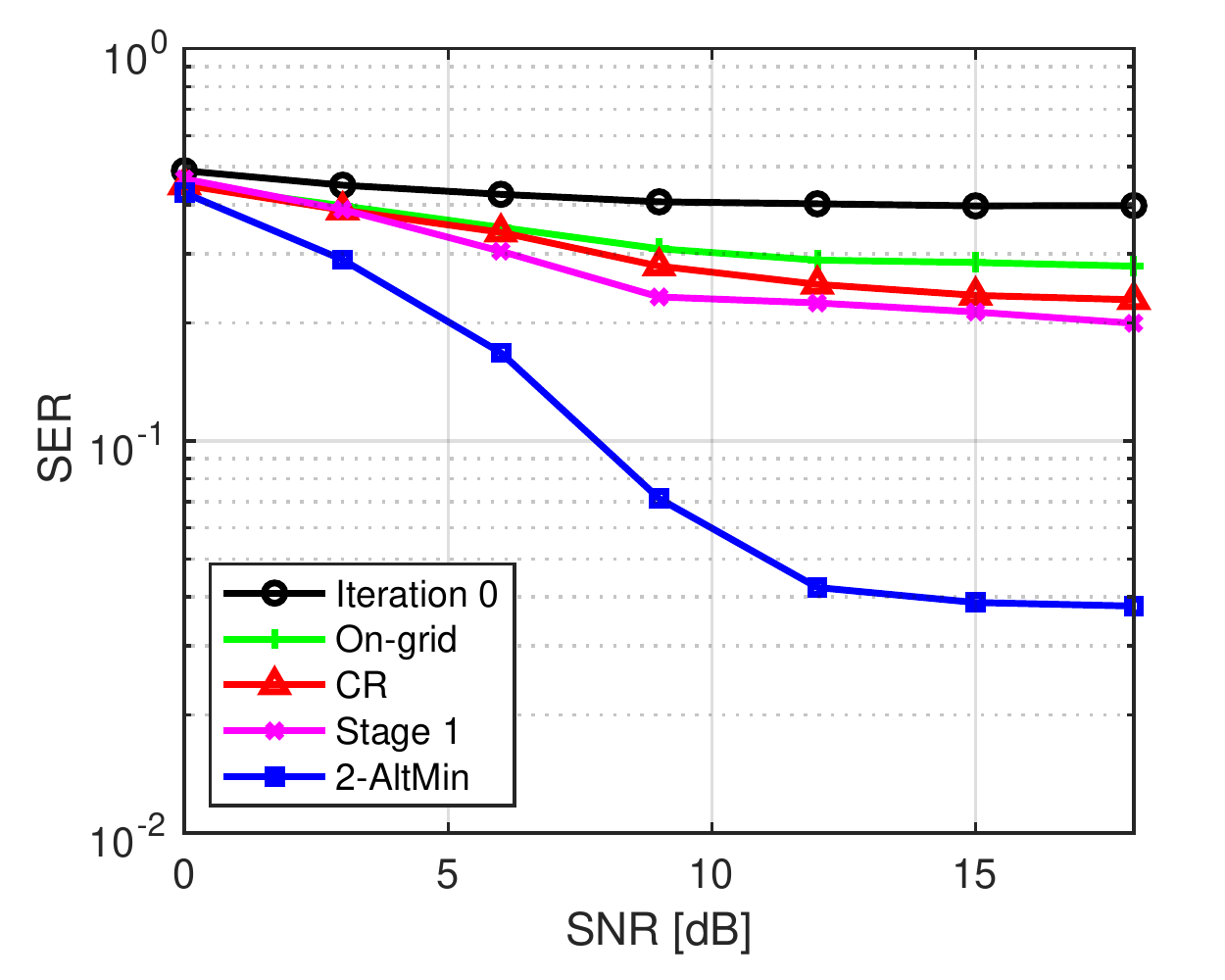}}
	\subfloat[][]{\includegraphics[width=3.2in]{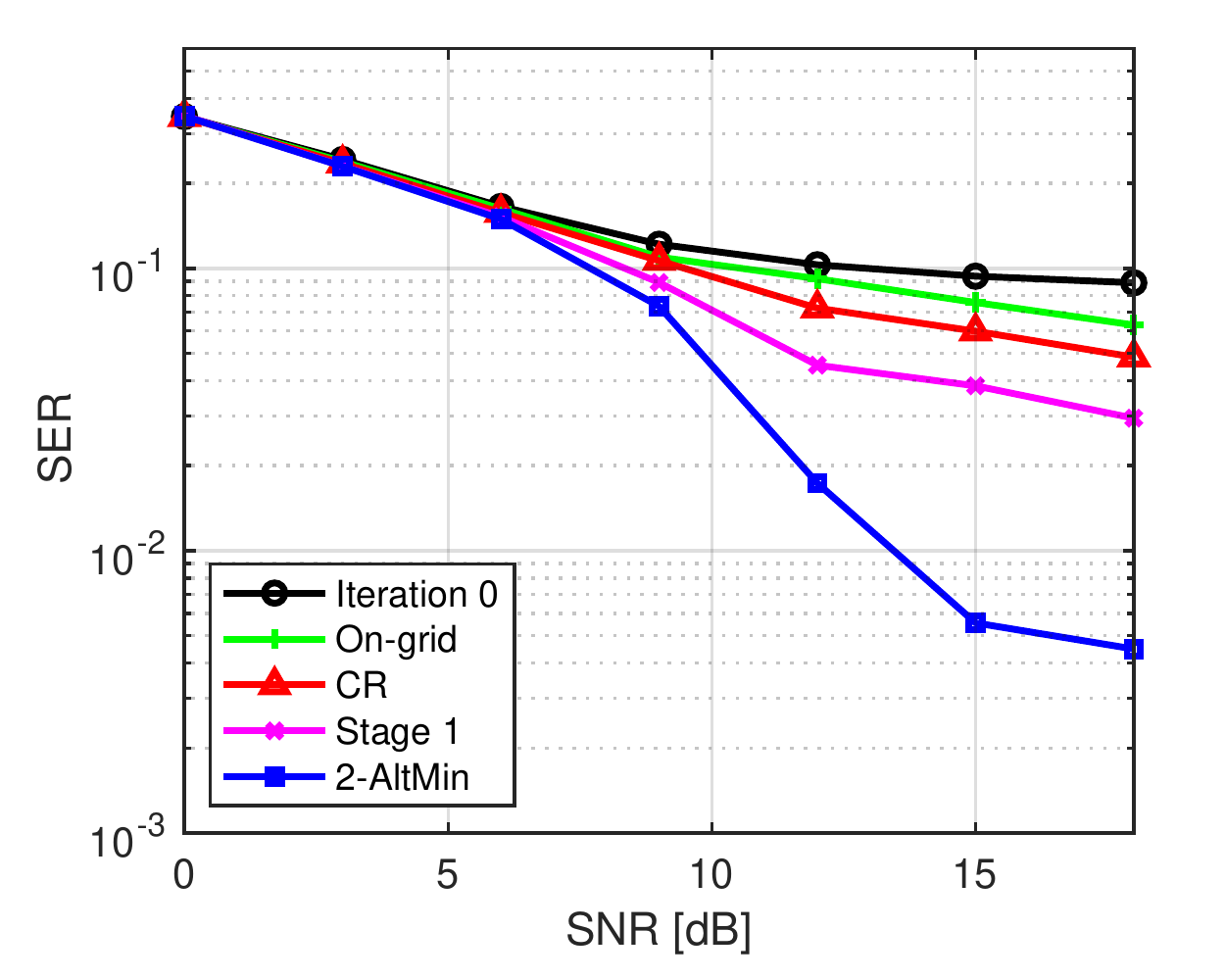}}

	\caption{SER performance comparison when the ISR of communication is (a) 5~dB and (b) -5~dB. }
	\label{fig:SER}
\end{figure*}

\begin{figure*}
	\centering
	
	\subfloat[][]{\includegraphics[width=3.2in]{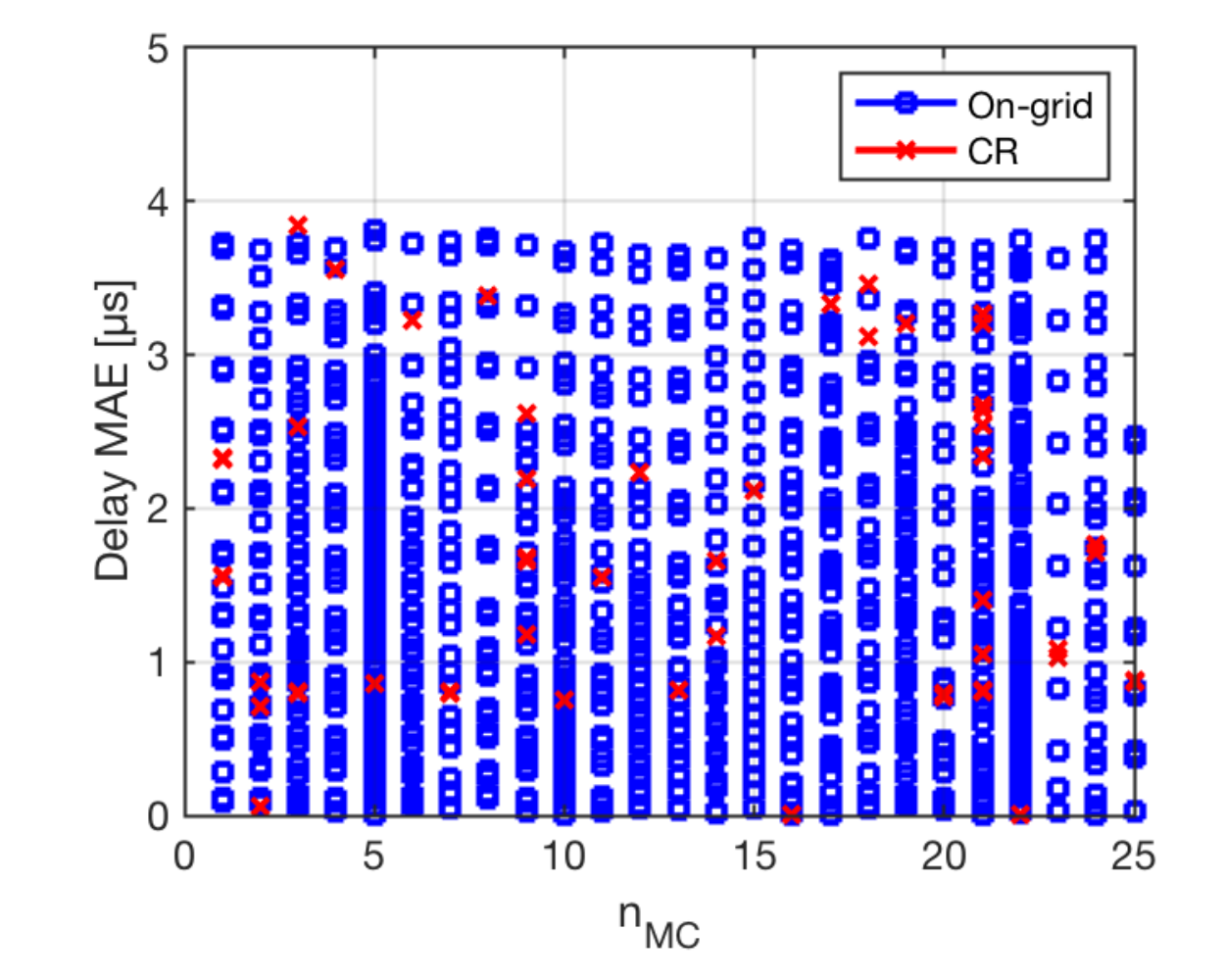}}
	\subfloat[][]{\includegraphics[width=3.2in]{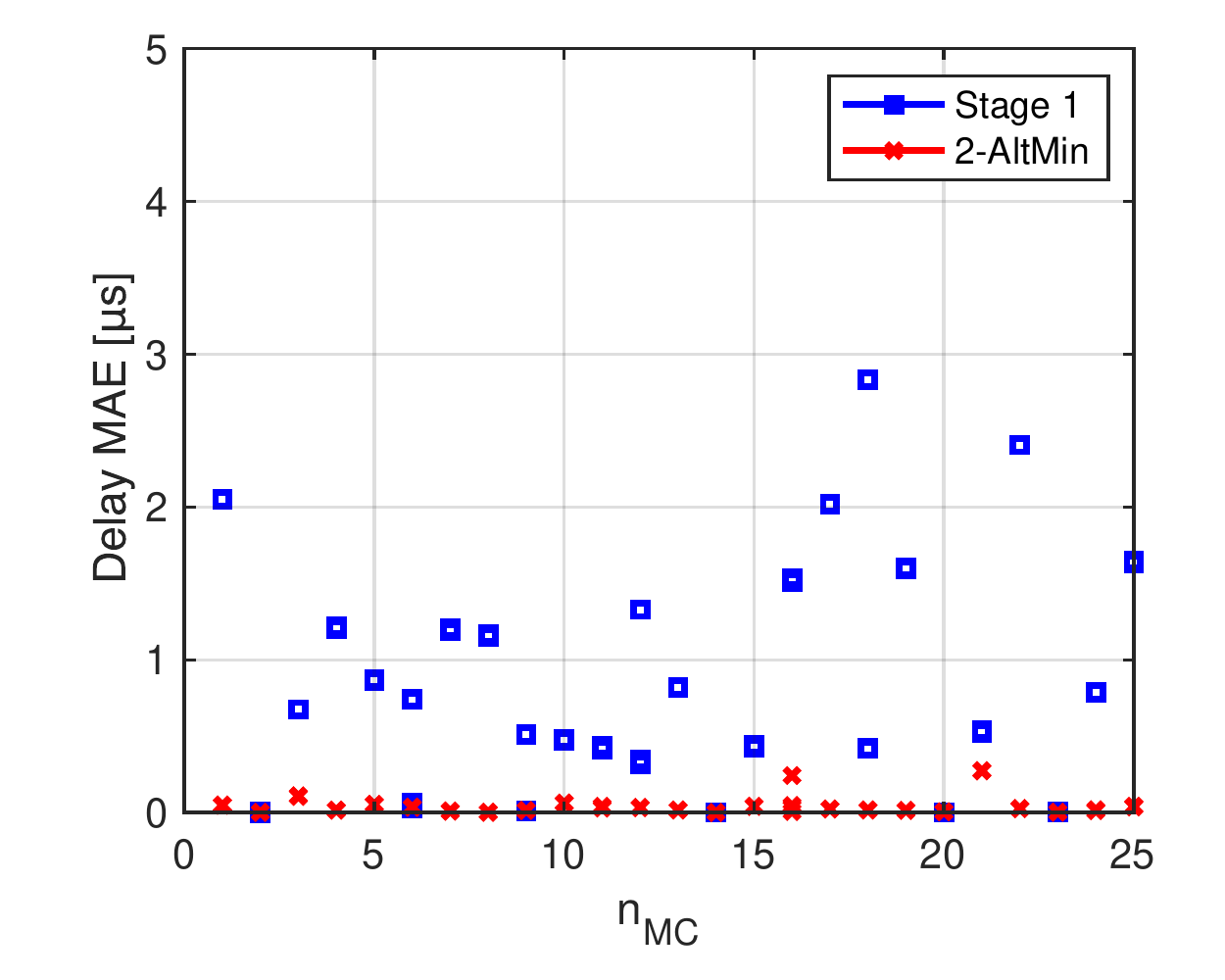}}
	
	\caption{Realizations of the delay minimum absolute errors. (a) On-grid and CR methods, (b) Stage 1 and 2-AltMin methods. }
	\label{fig:distribution}
\end{figure*}

In the first simulation, the number of scatterers is set as $M_r=2$ and the length of radar pulse is set at $L=10$. Fig.~\ref{fig:example} gives the signal at the communication RX and the interference and data for demodulation when the ISR is set at $-5~\text{dB}$. We can find that the effect of interference is significant even if there are only two multi-paths radar echo and the $\text{ISR}$ is moderate. Then, we compare the SER performance of various algorithms. In Fig.~\ref{fig:SER}, the effect of the SNR is studied: the on-grid, CR and 2-AltMin methods all provide better SER performance than Iteration 0. The 2-AltMin method also outperforms the on-grid and CR methods in all situations. It is worth mentioning that there is a significant improvement when the Stage 2 of 2-AltMin is used in all cases.

\begin{figure*}
	\centering
	
	\subfloat[][]{\includegraphics[width=3.2in]{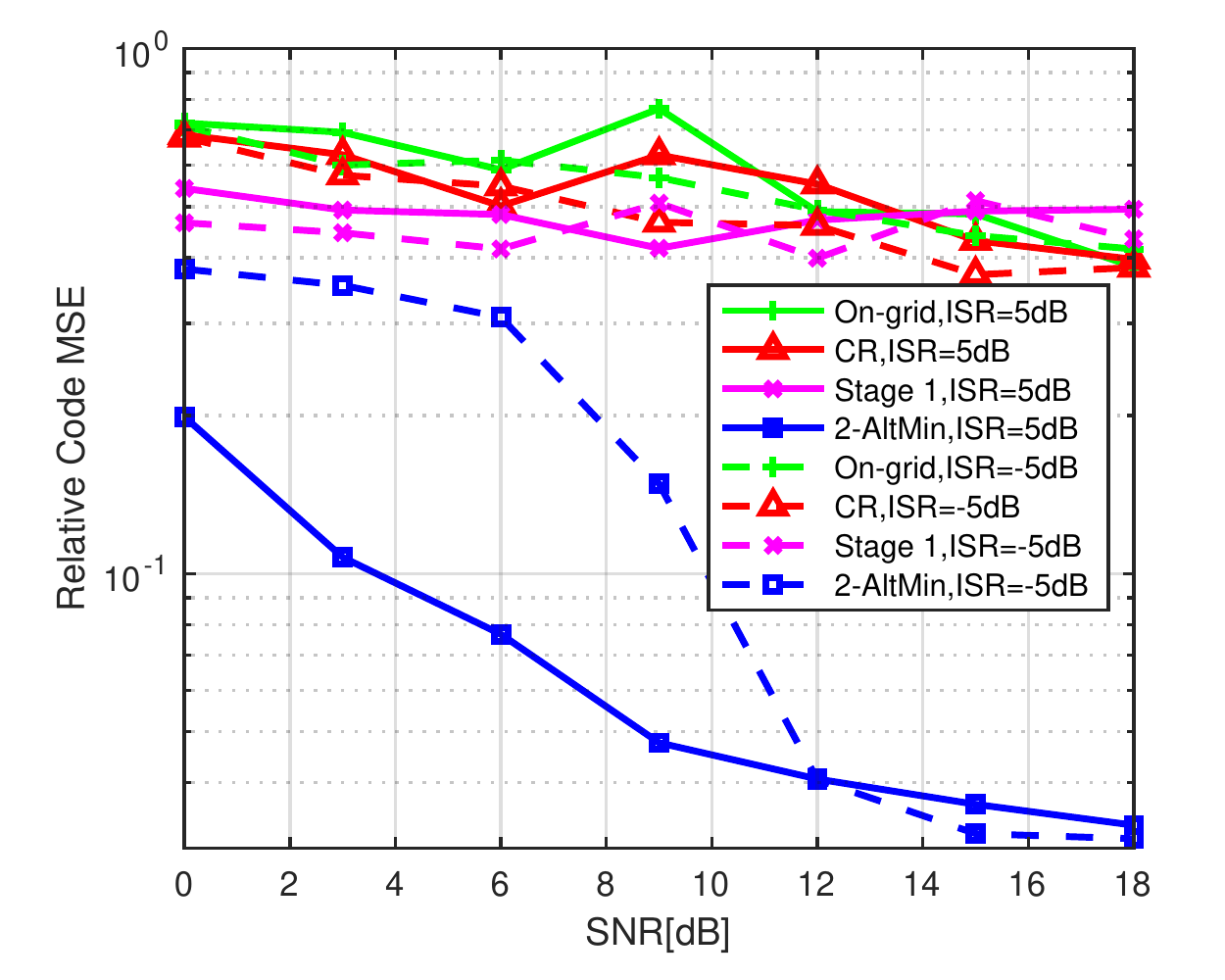}}
	\subfloat[][]{\includegraphics[width=3.2in]{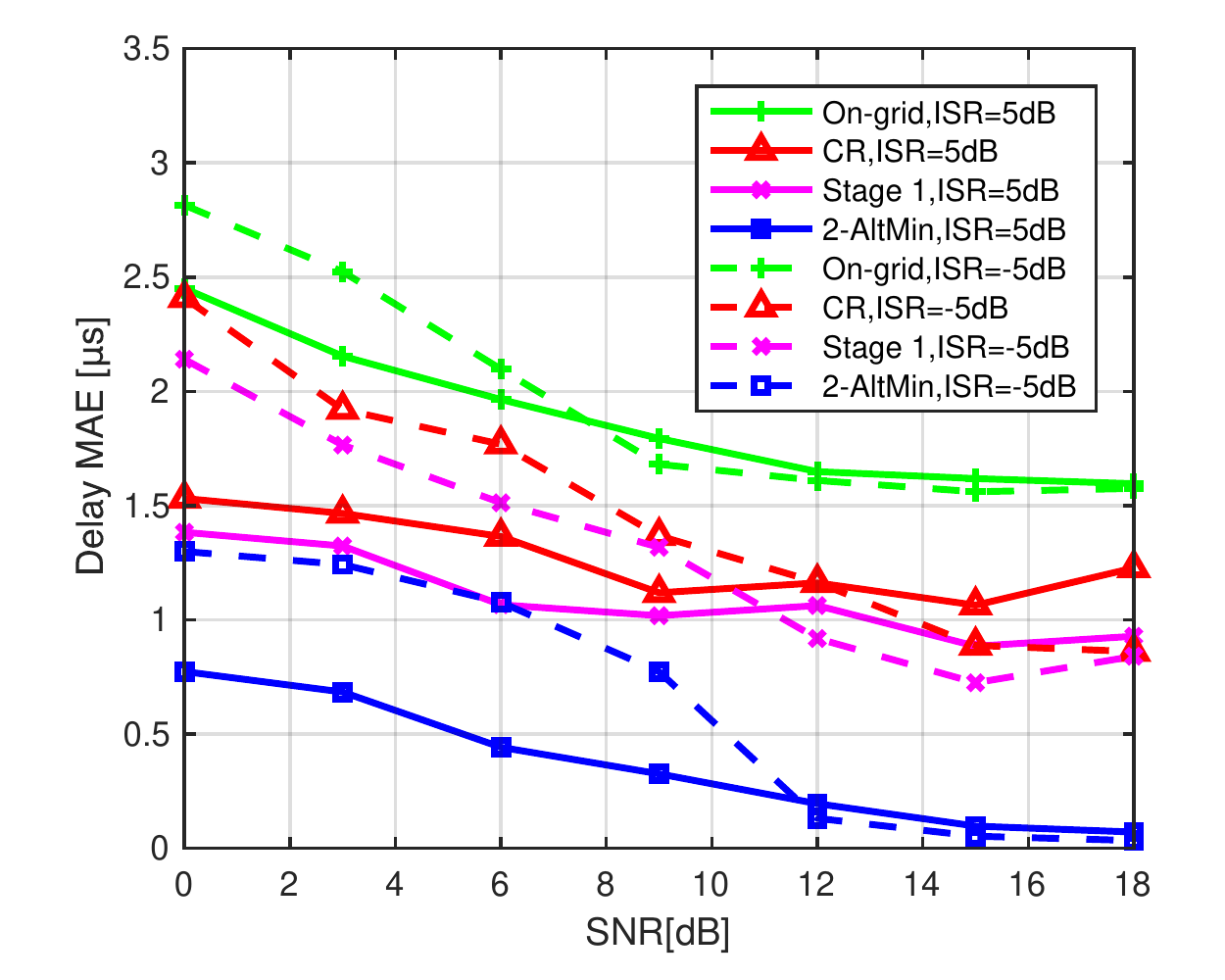}}
	
	\caption{(a) Relative waveform MSE performance comparisons. (b) Delay MAE performance comparisons. }
	\label{fig:Accuracy}
\end{figure*}

We then evaluate the relative code MSE and delay MAE of the proposed methods. We first plot the delay minimum absolute errors $\text{AE}_{\ell}^{(n_{\text{MC}})}$ of different methods when the SNR is 15~dB in Fig.~\ref{fig:distribution}. We can clearly see that the on-grid, CR and Stage 1 methods all reach many local optima, while the 2-AltMin method reaches the global optimum with high probability. And the on-grid method produces a large number of delay false alarms, because some columns of $\bm \Upsilon$ in \eqref{eq:Upsilon} are {\em coherent}. Then, the relative waveform MSE and delay MAE are plotted against the SNR in Fig.~\ref{fig:Accuracy}. Note that when the SNR is low, there are some very large delay minimum absolute errors, which affect the analysis of the average. Hence we remove the minimum absolute errors that are larger than $5~\mu s$. Then the delay MAE in Fig.~\ref{fig:Accuracy}(b) is calculated according to \eqref{eq:MAE}. As expected, the 2-AltMin method provides much better accuracy than other methods in all situations. In addition, we can see that the interference estimation accuracy may not necessarily improve with the ISR. When the SNR is low, the delay estimation performance is better when $\text{ISR} = 5~\text{dB}$, because strong radar interference can prevail the noise. While when the SNR is large, the delay estimation performance is better when $\text{ISR} = -5~\text{dB}$ because large SNRs guarantee good demodulation performance, with a beneficial effect on the radar interference estimation due to the coupling.

\begin{figure*}
	\centering
	
	\subfloat[][]{\includegraphics[width=3.2in]{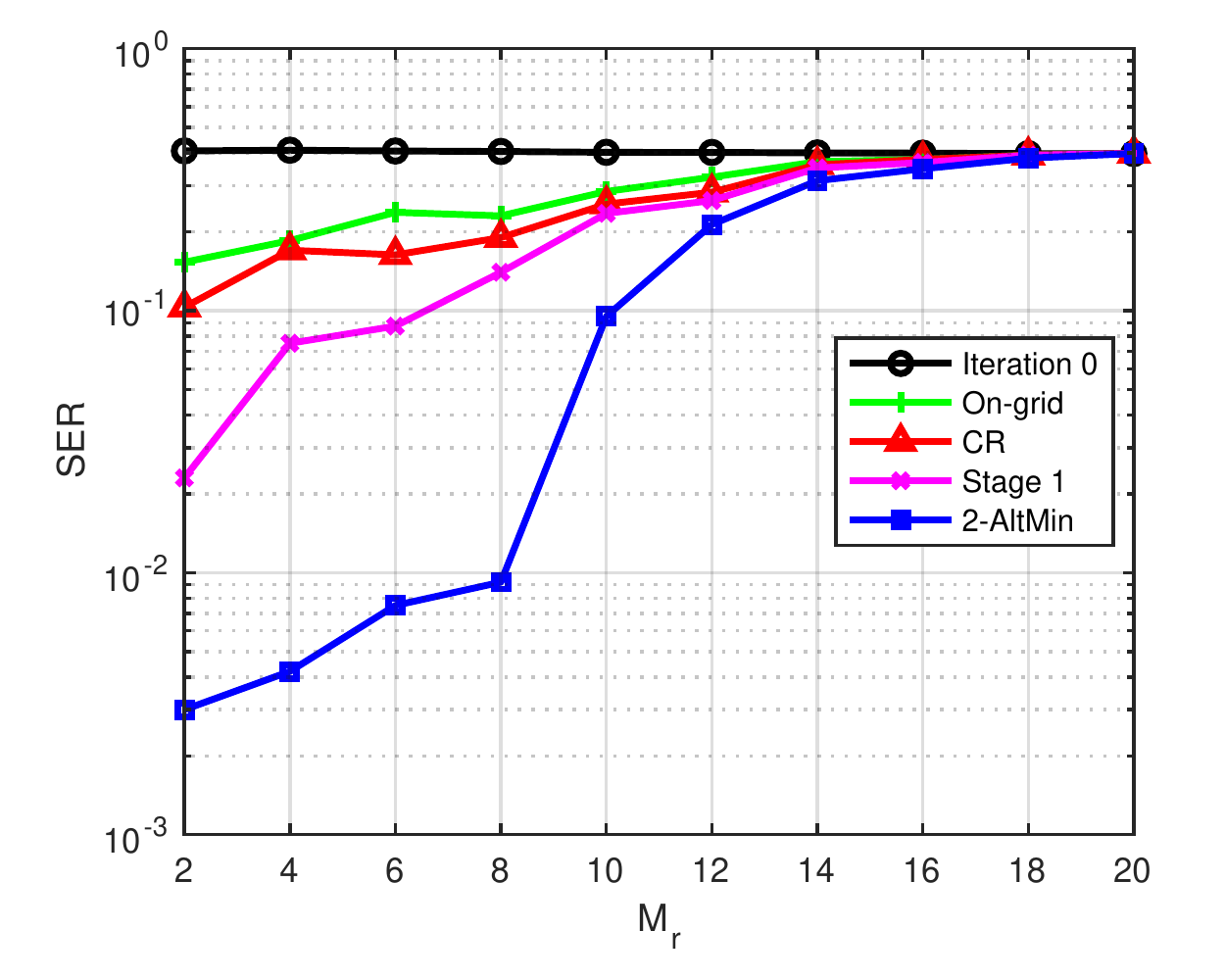}}
	\subfloat[][]{\includegraphics[width=3.2in]{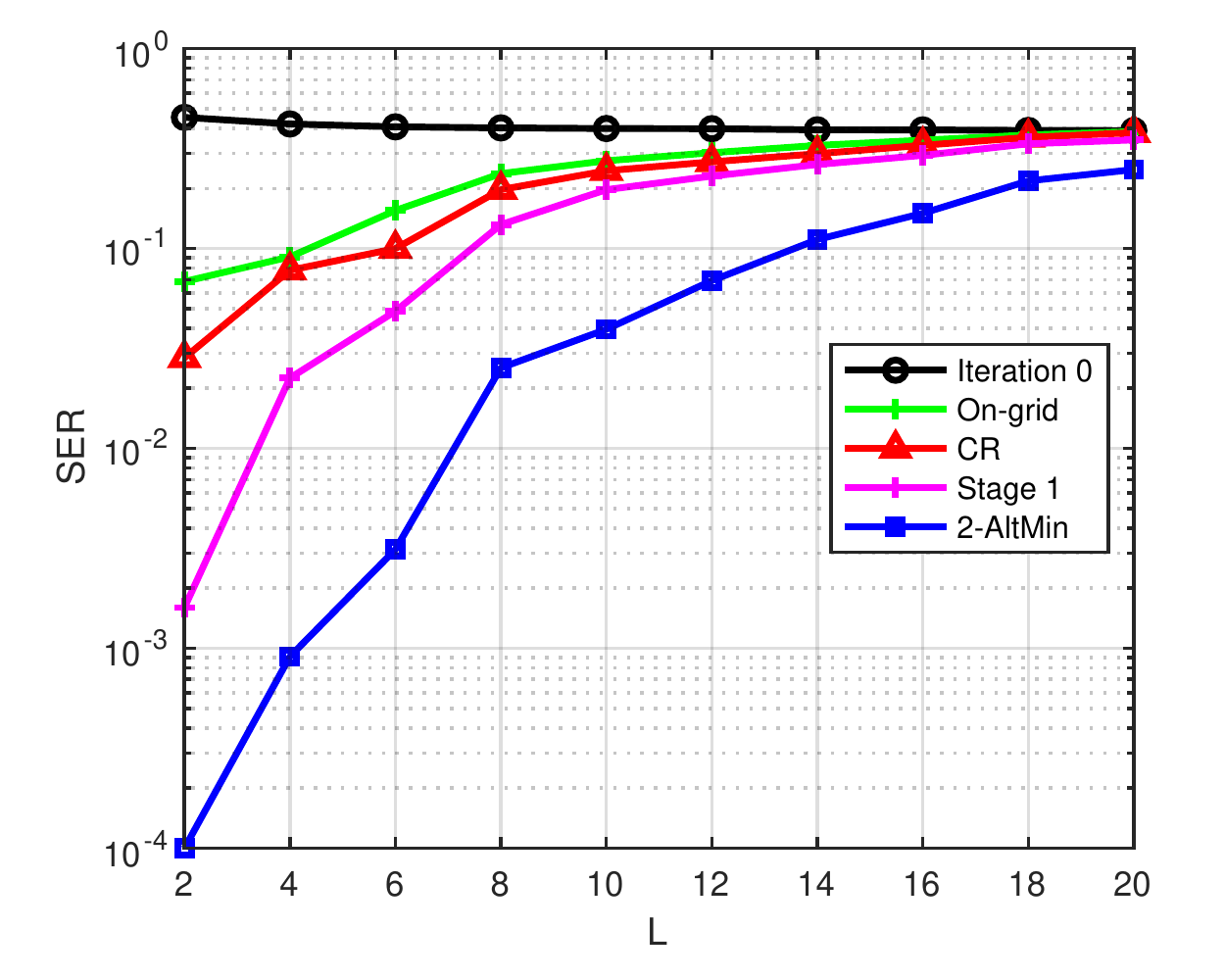}}
	
	\caption{SER performance against (a) $M_r$, and (b) $L$.}
	\label{fig:varing}
\end{figure*}

The effects of $M_r$ and $L$ are shown in Fig.~\ref{fig:varing}. The simulations are run with an SNR of 15~dB and an ISR of 5~dB. In Fig.~\ref{fig:varing}(a), we set $L=6$ and plot the SER against the number of scatterers: As $M_r$ increases, the sparsity of the problem is reduced, and the sources of interference - with the respective unknown parameters to be estimated - increase, which obviously results in a visible performance degradation for all algorithms. In Fig.~\ref{fig:varing}(b), the number of scatterers is set as $M_r=2$ and we examine the SER behavior for varying radar pluse length $L$. A performance degradation is also evident for all algorithms.

\begin{figure}[!htp]
	\centering
	
	\subfloat{\includegraphics[width=3.2in]{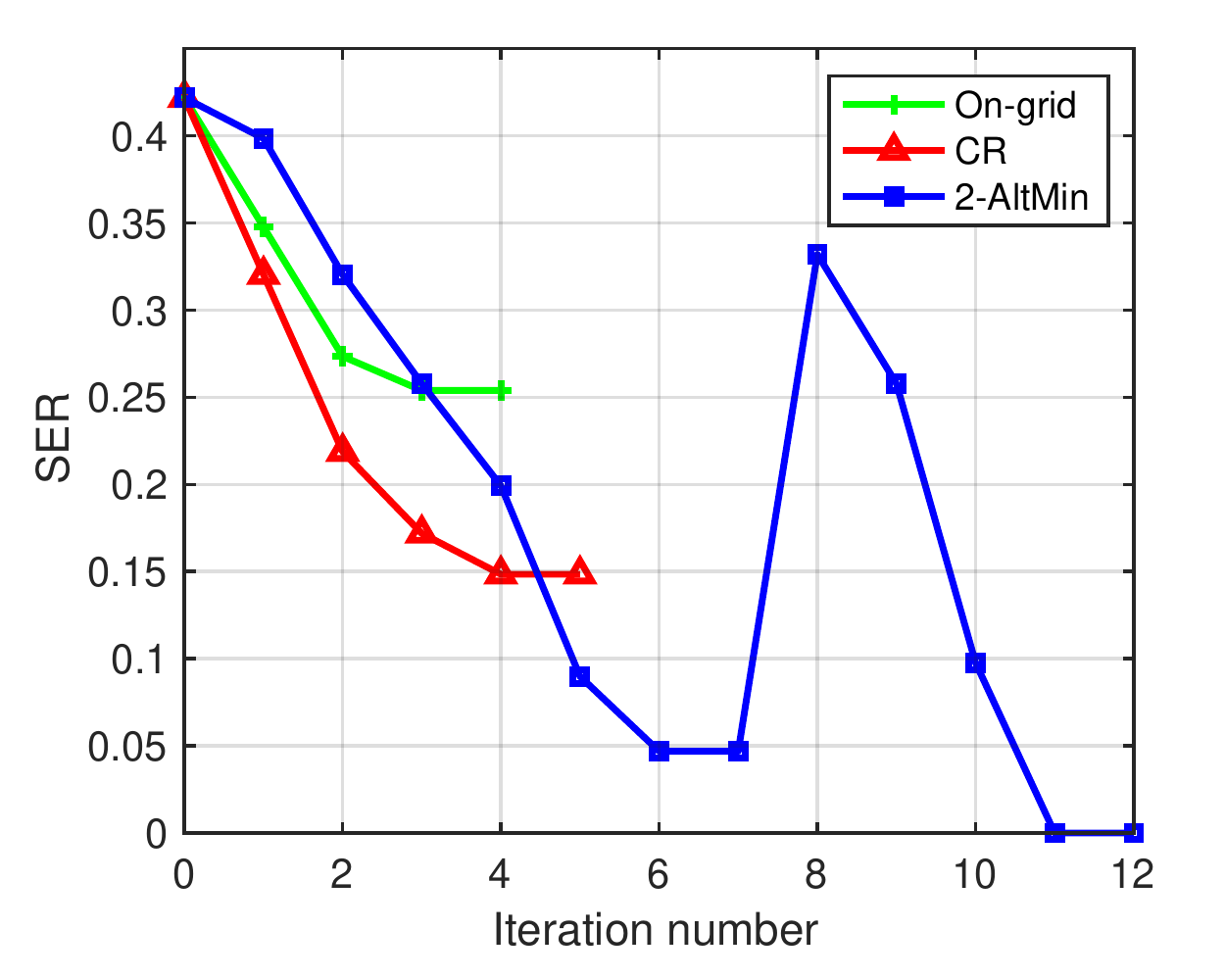}}
	
	\caption{Convergence behavior of different algorithms.}
	\label{fig:speed}
\end{figure}

Finally, we give an example of the convergence behavior of the three algorithms, which is shown in Fig.~\ref{fig:speed}. The number of scatterers, the length of radar pulse, the ISR and the SNR are respectively set as $M_r=2$, $L=10$, $\text{ISR} = 5~\text{dB}$ and $\text{SNR} = 15~\text{dB}$. The on-grid method takes 1616.3 seconds with 4 iterations by using CVX~\cite{boyd2004convex}. The CR method takes 1203.5 seconds with 5 iterations by using CVX, while the 2-AltMin method only takes 2.3 seconds with 12 iterations total (8 iterations in Stage 1 and 4 iterations in Stage 2). The experiments were carried out on a MacBook Pro computer with a 2.3 GHz Intel Core i5 CPU and 8 GB of RAM. The proposed 2-AltMin method is substantially faster than the CR method and the on-grid method and appears much well suited for real-time implementations.

\section{Conclusions}

In this paper, we have proposed two algorithms for removing the radar interference to facilitate more reliable data demodulation in a communication system overlaid with a radar system. The first one is based on forcing an atomic norm constraint, and estimating the combination of the radar parameters by solving a convex problem under some relaxations. The second algorithm estimates the radar parameters and the communication demodulation errors by two-stage processing. The first stage obtains a local optimum by alternating minimization, and the second stage infers the global optimum in a higher dimensional space by using the estimates of the first stage. The atomic norm and the $\ell_0$-norm are used to exploit the sparsity of the radar signal components and the sparsity of the demodulation error, respectively. Simulation results show that both algorithms provide much better SER performance compared to the conventional on-grid method. Moreover, the proposed 2-AltMin algorithm offers superior performance and is computationally efficient.

\appendix

\subsection{Proof of \eqref{eq:cost-tau}}

By noting that $(\bm \Phi^{-1})^H (\bm \Phi^{-1}) \succeq 0$ we have
\begin{align}
\label {eq:proof}
\arg\max_{\tau \in [0,1)}|\langle \bm \Phi \bm a({\tau}), \bm r_{\text{res}} \rangle|  =&~ {\arg\max_{\tau \in [0,1)}} { | \bm r_{\text{res}}^H  \bm \Phi \bm a({\tau}) |  } \nonumber \\
=&~ {\arg\max_{\tau \in [0,1)}} { | \bm r_{\text{res}}^H (\bm \Phi^{-1})^H (\bm \Phi^{-1}) \bm \Phi \bm a({\tau})  | } \nonumber \\
=&~ {\arg\max_{\tau \in [0,1)}} { | (\bm \Phi^{-1} \bm r_{\text{res}})^H  \bm a({\tau}) |  } \nonumber \\
=&~ {\arg\max_{\tau \in [0,1)}} {  ((\bm \Phi^{-1} \bm r_{\text{res}})^H  \bm a({\tau}) )^H (\bm \Phi^{-1} \bm r_{\text{res}})^H  \bm a({\tau})   } \nonumber \\
=&~ {\arg\max_{\tau \in [0,1)}} { \text{Tr}\{ \bm a({\tau})\bm a({\tau})^H (\bm \Phi^{-1} \bm r_{\text{res}}) (\bm \Phi^{-1} \bm r_{\text{res}})^H \}  } \nonumber \\
=&~ {\arg\min_{\tau \in [0,1)}} { \text{Tr}\{ \underbrace{( \bm I_{N_d}  - \frac{\bm a({\tau})\bm a({\tau})^H}{N_d} )}_{\bm {A}^{\perp}(\tau)}  \underbrace{(\bm \Phi^{-1} \bm r_{\text{res}}) (\bm \Phi^{-1} \bm r_{\text{res}})^H}_{ \bm{R}_{\text{res}}} \}  }.
\end{align}
Since $\bm a(\tau) = [1, e^{i2 \pi \tau},  ...,e^{i2\pi (N_d - 1) \tau}]^T$, then $\frac{1}{N_d} \bm a({\tau})^H = (\bm a({\tau})^H \bm a({\tau}))^{-1}\bm a({\tau})^H = \bm a({\tau})^{ \dagger}$. Thus we have ${\bm {A}^{\perp}(\tau)} = \bm I_{N_d}  - \bm a({\tau})\bm a({\tau})^{ \dagger}$ in the last line.

\begin{algorithm}[!bp]\small
	\label{tab:A3}
	\caption{Backtracking line search}
	\begin{tabular}{lcl}
	        Input ${\mathcal{L}}(\tau)$, ${\tau}^{i}$, ${\mathcal{D}}(\tau)$, ${\rho} \in (0,1)$ and $\bar{\rho} \in (0,1/2)$.\\
	        1, Initialize $\mu_i=1$.\\
	        2, \sf{Repeat} \\
	        3, \hspace{0.4cm} $\mu_i$ = ${\rho} \mu_i$ \\
	        4, \sf{Until} $ {\mathcal{L}}({\tau}^{i} - \mu_i {\mathcal{D}}(\tau^i)) \leq  {\mathcal{L}}({\tau}^{i}) - \bar{\rho} \mu_i  \| {\mathcal{D}}(\tau^i)  \|_2^2$. \\
	        \midrule
	        \sf{Return} $\mu_i$.
	\end{tabular}
\end{algorithm}

\subsection{Backtracking Line Search}
The backtracking line search approach ensures the selected step size is small enough to guarantee a sufficient decrease of the cost function but not too small. For simplify, define the objective functions for \eqref{eq:tau00}, \eqref{eq:tau0} and \eqref{eq:g1} respectively as ${\mathcal{L}}(\tau) = \text{Tr}\{ \bm {A}^{\perp}(\tau) \bm{R}_{\text{res}} \}$, ${\mathcal{L}}(\bm \tau) =  \text{Tr}\{\bm {P}^{\perp}(\bm{\tau}) \bm {R}\} $ and ${\mathcal{L}}(\bm g) =  \left \| \bm{\bar z} - \bm W \bm g \right \|_2^2 $. And define their search directions respectively as ${\mathcal{D}}(\tau) = K(\tau)^{-1} {p(\tau)}$, ${\mathcal{D}}(\bm\tau) = \bm K(\bm\tau)^{-1} {\bm p(\bm\tau)}$ and ${\mathcal{D}}(\bm g) = -\bm q_{\text{C}}(\bm g)$. As an example, in Algorithm 2 we summarize the backtracking line search for calculating $\mu_i$ in \eqref{eq:tau00}. Then $\bar\mu_i$ in \eqref{eq:tau0} and $\tilde\mu_i$ in \eqref{eq:g1} can be obtained with Algorithm 2 by replacing $({\mathcal{L}}(\tau),{\mathcal{D}}(\tau),{\tau}^{i},\mu_i)$ with $({\mathcal{L}}(\bm\tau),{\mathcal{D}}(\bm\tau),{\bm\tau}^{i},\bar\mu_i)$ and $({\mathcal{L}}(\bm g),{\mathcal{D}}(\bm g),{\bm g}^{i},\tilde\mu_i)$, respectively.

\bibliographystyle{IEEEtran}
\bibliography{database} 

\begin{thebibliography}{10}
\providecommand{\url}[1]{#1}
\csname url@samestyle\endcsname
\providecommand{\newblock}{\relax}
\providecommand{\bibinfo}[2]{#2}
\providecommand{\BIBentrySTDinterwordspacing}{\spaceskip=0pt\relax}
\providecommand{\BIBentryALTinterwordstretchfactor}{4}
\providecommand{\BIBentryALTinterwordspacing}{\spaceskip=\fontdimen2\font plus
\BIBentryALTinterwordstretchfactor\fontdimen3\font minus
  \fontdimen4\font\relax}
\providecommand{\BIBforeignlanguage}[2]{{%
\expandafter\ifx\csname l@#1\endcsname\relax
\typeout{** WARNING: IEEEtran.bst: No hyphenation pattern has been}%
\typeout{** loaded for the language `#1'. Using the pattern for}%
\typeout{** the default language instead.}%
\else
\language=\csname l@#1\endcsname
\fi
#2}}
\providecommand{\BIBdecl}{\relax}
\BIBdecl

\bibitem{griffiths2015radar}
H.~Griffiths, L.~Cohen, S.~Watts, E.~Mokole, C.~Baker, M.~Wicks, and S.~Blunt,
  ``Radar spectrum engineering and management: technical and regulatory
  issues,'' \emph{Proc. IEEE}, vol. 103, no.~1, pp. 85--102, Jan. 2015.

\bibitem{hassanien2016dual}
A.~Hassanien, M.~G. Amin, Y.~D. Zhang, and F.~Ahmad, ``{Dual-Function
  Radar-Communications: Information Embedding Using Sidelobe Control and
  Waveform Diversity.}'' \emph{IEEE Trans. Sig. Proc}, vol.~64, no.~8, pp.
  2168--2181, 2016.

\bibitem{Heath}
P.~Kumari, N.~Gonzalez-Prelcic, and R.~W. Heath, ``Investigating the ieee
  802.11ad standard for millimeter wave automotive radar,'' in \emph{Vehicular
  Technology Conference (VTC Fall), 2015 IEEE 82nd}.\hskip 1em plus 0.5em minus
  0.4em\relax IEEE, 2015, pp. 1--5.

\bibitem{802.11ad}
E.~Grossi, M.~Lops, L.~Venturino, and A.~Zappone, ``Opportunistic radar in
  802.11ad networks,'' \emph{IEEE Transactions on Signal Processing}, vol.~66,
  no.~9, pp. 2441--2454, 2018.

\bibitem{chiriyath2016inner}
A.~R. Chiriyath, B.~Paul, G.~M. Jacyna, and D.~W. Bliss, ``Inner bounds on
  performance of radar and communications co-existence,'' \emph{IEEE Trans.
  Signal Process.}, vol.~64, no.~2, pp. 464--474, Jan. 2016.

\bibitem{hessar2016spectrum}
F.~Hessar and S.~Roy, ``Spectrum sharing between a surveillance radar and
  secondary {Wi-Fi} networks,'' \emph{IEEE Trans. Aerosp. Electron. Syst.},
  vol.~52, no.~3, pp. 1434--1448, Jun. 2016.

\bibitem{ding2016modified}
Z.~Ding, B.~Shu, W.~Yin, T.~Zeng, and T.~Long, ``A modified frequency domain
  algorithm based on optimal azimuth quadratic factor compensation for
  geosynchronous {SAR} imaging,'' \emph{IEEE J. Sel. Topics Appl. Earth
  Observ.}, vol.~9, no.~3, pp. 1119--1131, 2016.

\bibitem{babaei2013practical}
A.~Babaei, W.~H. Tranter, and T.~Bose, ``A practical precoding approach for
  radar/communications spectrum sharing,'' in \emph{Proc. 8th Int. Conf.
  Cognitive Radio Oriented Wireless Netw.}, 2013, pp. 13--18.

\bibitem{sodagari2012projection}
S.~Sodagari, A.~Khawar, T.~C. Clancy, and R.~McGwier, ``A projection based
  approach for radar and telecommunication systems coexistence,'' in
  \emph{Proc. Global Commun. Conf.}, 2012, pp. 5010--5014.

\bibitem{deng2013interference}
H.~Deng and B.~Himed, ``Interference mitigation processing for spectrum-sharing
  between radar and wireless communications systems,'' \emph{IEEE Trans.
  Aerosp. Electron. Syst.}, vol.~49, no.~3, pp. 1911--1919, 2013.

\bibitem{aubry2014radar}
A.~Aubry, A.~De~Maio, M.~Piezzo, and A.~Farina, ``Radar waveform design in a
  spectrally crowded environment via nonconvex quadratic optimization,''
  \emph{IEEE Trans. Aerosp. Electron. Syst.}, vol.~50, no.~2, pp. 1138--1152,
  2014.

\bibitem{huang2015radar}
K.-W. Huang, M.~Bic{\u{a}}, U.~Mitra, and V.~Koivunen, ``Radar waveform design
  in spectrum sharing environment: Coexistence and cognition,'' in \emph{Proc.
  Radar Conf.}, 2015, pp. 1698--1703.

\bibitem{lioptimum}
B.~Li, A.~Petropulu, and W.~Trappe, ``Optimum co-design for spectrum sharing
  between matrix completion based {MIMO} radars and a {MIMO} communication
  system,'' \emph{IEEE Trans. Signal Process.}, vol.~64, no.~17, pp.
  4562--4575, 2016.

\bibitem{zheng2018joint}
L.~Zheng, M.~Lops, X.~Wang, and E.~Grossi, ``Joint design of overlaid
  communication systems and pulsed radars,'' \emph{IEEE Trans. Signal
  Process.}, vol.~66, no.~1, pp. 139--154, 2018.

\bibitem{turlapaty2014joint}
A.~Turlapaty and Y.~Jin, ``A joint design of transmit waveforms for radar and
  communications systems in coexistence,'' in \emph{Proc. Radar Conf.}, 2014,
  pp. 0315--0319.

\bibitem{khawar2014spectrum}
A.~Khawar, A.~Abdel-Hadi, and T.~C. Clancy, ``Spectrum sharing between {S}-band
  radar and {LTE} cellular system: A spatial approach,'' in \emph{Proc. Int.
  Symp. Dyn. Spectrum Access Netw.}, 2014, pp. 7--14.

\bibitem{manolakos2012blind}
A.~Manolakos, Y.~Noam, K.~Dimou, and A.~J. Goldsmith, ``Blind null-space
  tracking for {MIMO} underlay cognitive radio networks,'' in \emph{Proc.
  Global Commun. Conf.}, 2012, pp. 1223--1229.

\bibitem{zheng2018adaptive}
L.~Zheng, M.~Lops, and X.~Wang, ``Adaptive interference removal for
  uncoordinated radar/communication coexistence,'' \emph{IEEE J. Sel. Top.
  Signal Proces.}, vol.~12, no.~1, pp. 45--60, 2018.

\bibitem{liu2014joint}
J.~Liu, H.~Li, and B.~Himed, ``Joint optimization of transmit and receive
  beamforming in active arrays,'' \emph{IEEE Signal Process. Lett.}, vol.~21,
  no.~1, pp. 39--42, Jan. 2014.

\bibitem{candes2011compressed}
E.~J. Candes, Y.~C. Eldar, D.~Needell, and P.~Randall, ``Compressed sensing
  with coherent and redundant dictionaries,'' \emph{Appl. Comput. Harmon.
  Anal.}, vol.~31, no.~1, pp. 59--73, 2011.

\bibitem{zhang2018recovery}
X.~Zhang, W.~Cui, and Y.~Liu, ``Recovery of structured signals with prior
  information via maximizing correlation,'' \emph{IEEE Trans. Signal Process.},
  vol.~66, no.~12, pp. 3296--3310, 2018.

\bibitem{chi2011sensitivity}
Y.~Chi, L.~L. Scharf, A.~Pezeshki, and A.~R. Calderbank, ``Sensitivity to basis
  mismatch in compressed sensing,'' \emph{IEEE Trans. Signal Process.},
  vol.~59, no.~5, pp. 2182--2195, 2011.

\bibitem{stankovic2013compressive}
L.~Stankovi{\'c}, I.~Orovi{\'c}, S.~Stankovi{\'c}, and M.~Amin, ``Compressive
  sensing based separation of nonstationary and stationary signals overlapping
  in time-frequency,'' \emph{IEEE Trans. Signal Process.}, vol.~61, no.~18, pp.
  4562--4572, 2013.

\bibitem{jokanovic2015reduced}
B.~Jokanovic and M.~Amin, ``Reduced interference sparse time-frequency
  distributions for compressed observations,'' \emph{IEEE Trans. Signal
  Process.}, vol.~63, no.~24, pp. 6698--6709, 2015.

\bibitem{studer2012recovery}
C.~Studer, P.~Kuppinger, G.~Pope, and H.~Bolcskei, ``Recovery of sparsely
  corrupted signals,'' \emph{IEEE Trans. Inf. Theory}, vol.~58, no.~5, pp.
  3115--3130, 2012.

\bibitem{candes2013super}
E.~J. Cand{\`e}s and C.~Fernandez-Granda, ``Super-resolution from noisy data,''
  \emph{J. Fourier Anal. Appl.}, vol.~19, no.~6, pp. 1229--1254, 2013.

\bibitem{candes2014towards}
E.~J. Cand\`es and C.~Fernandez-Granda, ``Towards a mathematical theory of
  super-resolution,'' \emph{Commun. Pure Appl. Math.}, vol.~67, no.~6, pp.
  906--956, 2014.

\bibitem{tang2013compressed}
G.~Tang, B.~N. Bhaskar, P.~Shah, and B.~Recht, ``Compressed sensing off the
  grid,'' \emph{IEEE Trans. Inf. Theory}, vol.~59, no.~11, pp. 7465--7490, Nov.
  2013.

\bibitem{bhaskar2013atomic}
B.~N. Bhaskar, G.~Tang, and B.~Recht, ``Atomic norm denoising with applications
  to line spectral estimation,'' \emph{IEEE Trans. Signal Process.}, vol.~61,
  no.~23, pp. 5987--5999, Dec. 2013.

\bibitem{tan2014direction}
Z.~Tan, Y.~C. Eldar, and A.~Nehorai, ``Direction of arrival estimation using
  co-prime arrays: A super resolution viewpoint,'' \emph{IEEE Trans. Signal
  Process.}, vol.~62, no.~21, pp. 5565--5576, Nov. 2014.

\bibitem{ling2017blind}
S.~Ling and T.~Strohmer, ``Blind deconvolution meets blind demixing: Algorithms
  and performance bounds,'' \emph{IEEE Trans. Inf. Theory}, vol.~63, no.~7, pp.
  4497--4520, 2017.

\bibitem{ahmed2014blind}
A.~Ahmed, B.~Recht, and J.~Romberg, ``Blind deconvolution using convex
  programming,'' \emph{IEEE Trans. Inf. Theory}, vol.~60, no.~3, pp.
  1711--1732, 2014.

\bibitem{chi2016guaranteed}
Y.~Chi, ``Guaranteed blind sparse spikes deconvolution via lifting and convex
  optimization,'' \emph{IEEE J. Sel. Top. Signal Proces.}, vol.~10, no.~4, pp.
  782--794, 2016.

\bibitem{zhang2017global}
Y.~Zhang, Y.~Lau, H.-w. Kuo, S.~Cheung, A.~Pasupathy, and J.~Wright, ``On the
  global geometry of sphere-constrained sparse blind deconvolution,'' in
  \emph{Proc. IEEE Conf. Comput. Vision Pattern Recogni.}, 2017, pp.
  4894--4902.

\bibitem{zhang2018structured}
Y.~Zhang, H.-w. Kuo, and J.~Wright, ``Structured local minima in sparse blind
  deconvolution,'' in \emph{Adv. Neural Inf. Process. Syst.}, 2018, pp.
  2324--2333.

\bibitem{engels2017advances}
F.~Engels, P.~Heidenreich, A.~M. Zoubir, F.~K. Jondral, and M.~Wintermantel,
  ``Advances in automotive radar: A framework on computationally efficient
  high-resolution frequency estimation,'' \emph{IEEE Signal Proc. Mag.},
  vol.~34, no.~2, pp. 36--46, 2017.

\bibitem{doufexi2002comparison}
A.~Doufexi, S.~Armour, M.~Butler, A.~Nix, D.~Bull, J.~McGeehan, and
  P.~Karlsson, ``A comparison of the hiperlan/2 and ieee 802.11 a wireless lan
  standards,'' \emph{IEEE Commun. mag.}, vol.~40, no.~5, pp. 172--180, 2002.

\bibitem{zhiguo2006moving}
W.~Zhiguo, L.~Xi, and F.~Yuanchun, ``Moving target position with through-wall
  radar,'' in \emph{Int. Conf. Radar}.\hskip 1em plus 0.5em minus 0.4em\relax
  IEEE, 2006, pp. 1--4.

\bibitem{wei2008detection}
G.~Wei, Y.~Zhou, and S.~Wu, ``Detection and localization of high speed moving
  targets using a short-range uwb impulse radar,'' in \emph{IEEE Radar
  Conf.}\hskip 1em plus 0.5em minus 0.4em\relax IEEE, 2008, pp. 1--4.

\bibitem{berger2010signal}
C.~R. Berger, B.~Demissie, J.~Heckenbach, P.~Willett, and S.~Zhou, ``Signal
  processing for passive radar using {OFDM} waveforms,'' \emph{IEEE J. Sel.
  Top. Signal Proces.}, vol.~4, no.~1, pp. 226--238, 2010.

\bibitem{molisch2012wireless}
A.~F. Molisch, \emph{Wireless {C}ommunications}.\hskip 1em plus 0.5em minus
  0.4em\relax John Wiley \& Sons, 2012.

\bibitem{hu2011efficient}
D.~Hu, L.~He, and X.~Wang, ``An efficient pilot design method for {OFDM}-based
  cognitive radio systems,'' \emph{IEEE Trans. Wireless Commun.}, vol.~10,
  no.~4, pp. 1252--1259, 2011.

\bibitem{yang2016super}
D.~Yang, G.~Tang, and M.~B. Wakin, ``Super-resolution of complex exponentials
  from modulations with unknown waveforms,'' \emph{IEEE Trans. Inf. Theory},
  vol.~62, no.~10, pp. 5809--5830, 2016.

\bibitem{yang2016enhancing}
Z.~Yang and L.~Xie, ``Enhancing sparsity and resolution via reweighted atomic
  norm minimization,'' \emph{IEEE Trans. Signal Process.}, vol.~64, no.~4, pp.
  995--1006, 2016.

\bibitem{li2018atomic}
S.~Li, D.~Yang, G.~Tang, and M.~Wakin, ``Atomic norm minimization for modal
  analysis from random and compressed samples,'' \emph{IEEE Trans. Signal
  Process.}, 2018.

\bibitem{boyd2004convex}
S.~Boyd and L.~Vandenberghe, \emph{Convex {O}ptimization}.\hskip 1em plus 0.5em
  minus 0.4em\relax Cambridge University Press, 2004.

\bibitem{naha2015determining}
A.~Naha, A.~K. Samanta, A.~Routray, and A.~K. Deb, ``Determining
  autocorrelation matrix size and sampling frequency for {MUSIC} algorithm,''
  \emph{IEEE Signal Process. Lett.}, vol.~22, no.~8, pp. 1016--1020, 2015.

\bibitem{tropp2007signal}
J.~A. Tropp and A.~C. Gilbert, ``Signal recovery from random measurements via
  orthogonal matching pursuit,'' \emph{IEEE Trans. Inf. Theory}, vol.~53,
  no.~12, pp. 4655--4666, 2007.

\bibitem{lagarias1998convergence}
J.~C. Lagarias, J.~A. Reeds, M.~H. Wright, and P.~E. Wright, ``Convergence
  properties of the {Nelder-Mead} simplex method in low dimensions,''
  \emph{SIAM Journal on Optimization}, vol.~9, no.~1, pp. 112--147, 1998.

\bibitem{fernandez2017demixing}
C.~Fernandez-Granda, G.~Tang, X.~Wang, and L.~Zheng, ``Demixing sines and
  spikes: Robust spectral super-resolution in the presence of outliers,''
  \emph{Information and Inference: A Journal of the IMA}, vol.~7, no.~1, pp.
  105--168, 2017.

\bibitem{bertsekas1999nonlinear}
D.~P. Bertsekas, \emph{Nonlinear {P}rogramming}.\hskip 1em plus 0.5em minus
  0.4em\relax Athena Scientific Belmont, 1999.

\bibitem{viberg1991detection}
M.~Viberg, B.~Ottersten, and T.~Kailath, ``Detection and estimation in sensor
  arrays using weighted subspace fitting,'' \emph{IEEE Trans. Signal Process.},
  vol.~39, no.~11, pp. 2436--2449, 1991.

\bibitem{viberg1991sensor}
M.~Viberg and B.~Ottersten, ``Sensor array processing based on subspace
  fitting,'' \emph{IEEE Trans. Signal Process.}, vol.~39, no.~5, pp.
  1110--1121, 1991.

\bibitem{chen2017low}
J.-C. Chen, ``{Low-PAPR} precoding design for massive multiuser {MIMO} systems
  via {Riemannian} manifold optimization,'' \emph{IEEE Commun. Lett.}, vol.~21,
  no.~4, pp. 945--948, 2017.

\bibitem{absil2009optimization}
P.-A. Absil, R.~Mahony, and R.~Sepulchre, \emph{Optimization {A}lgorithms on
  {M}atrix {M}anifolds}.\hskip 1em plus 0.5em minus 0.4em\relax Princeton
  University Press, 2009.

\bibitem{li2012isar}
G.~Li, H.~Zhang, X.~Wang, and X.-G. Xia, ``{ISAR} {2-D} imaging of uniformly
  rotating targets via matching pursuit,'' \emph{IEEE Trans. Aerosp. Electron.
  Syst.}, vol.~48, no.~2, pp. 1838--1846, 2012.

\end{thebibliography}

\end{document}